\documentclass[journal]{vgtc}                     

\onlineid{0}

\vgtccategory{Research}

\vgtcpapertype{system}

\newcommand{\thetitle}{ReVISit 2: A Full Experiment Life Cycle User Study Framework}

\title{\thetitle}

\author{%
  \authororcid{Zach Cutler}{0000-0002-2656-3413},
  \authororcid{Jack Wilburn}{0000-0002-7672-0798}, 
  \authororcid{Hilson Shrestha}{0009-0006-6603-6606},
  \authororcid{Yiren Ding}{0000-0001-8983-9117},
  \authororcid{Brian Bollen}{0000-0002-9557-1783},
  \\
  \authororcid{Khandaker Abrar Nadib}{0009-0006-7940-501X}, 
  \authororcid{Tingying He}{0000-0002-9670-5587},
  \authororcid{Andrew McNutt}{0000-0001-8255-4258},
  \authororcid{Lane Harrison}{0000-0003-3029-2799}, and
  \authororcid{Alexander Lex}{0000-0001-6930-5468}
}

\authorfooter{

  \item Zach Cutler, Jack Wilburn, Brian Bollen, Khandaker Abrar Nadib, Tingying He, Andrew McNutt, and Alexander Lex are with the University of Utah.
    \item Hilson Shrestha, Yiren Ding, and Lane Harrison are with WPI.
}

\abstract{%
Online user studies of visualizations, visual encodings, and interaction techniques are ubiquitous in visualization research.  
Yet, designing, conducting, and analyzing studies effectively is still a major burden.
Although various packages support such user studies, most solutions address only facets of the experiment life cycle, make reproducibility difficult, or do not cater to nuanced study designs or interactions.
We introduce reVISit 2, a software framework that supports visualization researchers at all stages of designing and conducting browser-based user studies.
ReVISit supports researchers in the design, debug \& pilot, data collection, analysis, and dissemination experiment phases by providing both technical affordances (such as replay of participant interactions) and sociotechnical aids (such as a mindfully maintained community of support).
It is a proven system that can be (and has been) used in publication-quality studies---which we demonstrate through a series of experimental replications.  
We reflect on the design of the system via interviews and an analysis of its technical dimensions.
Through this work, we seek to elevate the ease with which studies are conducted, improve the reproducibility of studies within our community, and support the construction of advanced interactive studies.
}


\teaser{
  \centering
  \includegraphics[width=\linewidth, alt={A view of a city with buildings peeking out of the clouds.}]{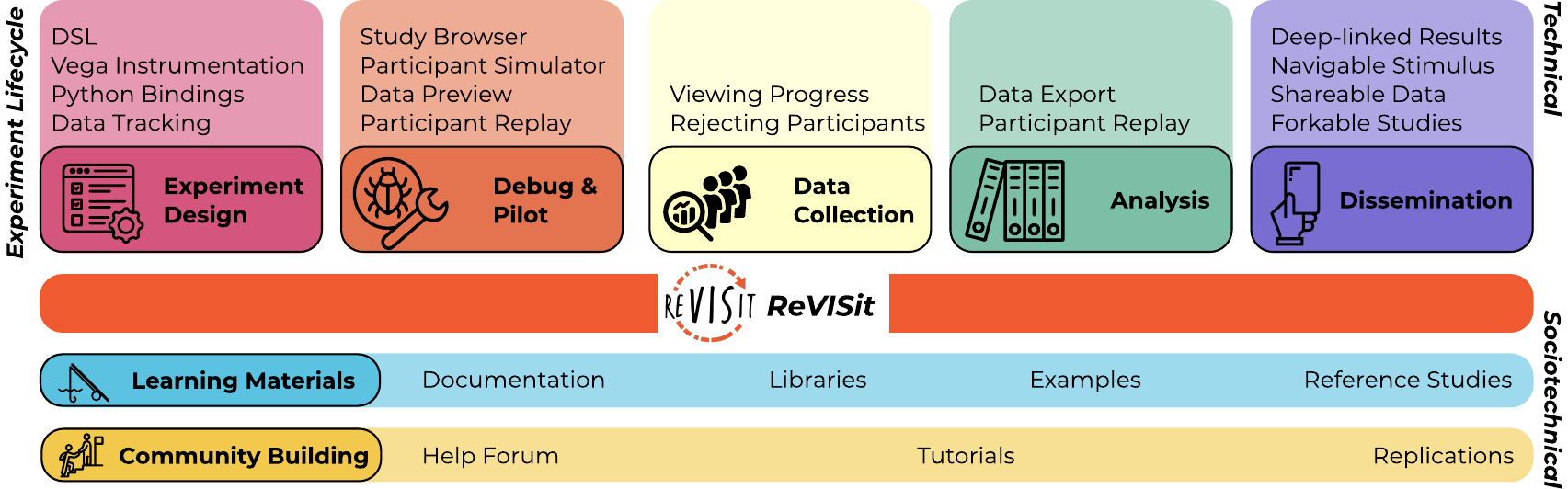}
  \caption{%
  	ReVISit 2 supports each stage of the online user study life cycle. In addition to this linear path, the experiment life cycle is populated by internal loops between stages. For instance, issues revealed in the piloting of an experiment can lead back to the design phase to address those issues, and the process of disseminating an experiment can beget ideas for new experiments or replications. 
  }
  \label{fig:teaser}
}

\graphicspath{{figs/}{figures/}{pictures/}{images/}{./}} 

\usepackage[svgnames, table]{xcolor}

\usepackage{tabu}                      
\usepackage{booktabs}                  
\usepackage{lipsum}                    
\usepackage{mwe}                       

\usepackage{mathptmx}                  
\usepackage{soul}
\usepackage[normalem]{ulem}
\usepackage{colortbl}
\usepackage{amsfonts}
\usepackage{adjustbox}
\usepackage{wrapfig}
\usepackage{cuted}
\usepackage{fancybox}
\usepackage{graphicx}
\usepackage{setspace}

\newcommand{\etal}{et al.}
\newcommand{\etals}{\etal{}'s}

\newcommand{\eg}{{e.g.,}}

\newcommand{\cf}{{cf.}}
\newcommand{\ala}{\`a la}

\newcommand{\figref}[1]{\hyperref[#1]{Fig.~\ref*{#1}}}
\newcommand{\appendixfigref}[1]{\hyperref[#1]{Appendix Fig.~\ref*{#1}}}
\newcommand{\secref}[1]{\hyperref[#1]{Sec.~\ref*{#1}}}

\newcommand{\parahead}[1]{\paraheadd{#1}.}
\newcommand{\paraheadd}[1]{%
    \vspace{0.5em}%
    \noindent%
    \textbf{\textit{#1}}%
}

\input ulem.sty\relax
\catcode`\@=11 %

\font\uwavefont=lasyb10 scaled 652
\def\uwave{%
  \bgroup
    \markoverwith{%
      \lower3.5\p@\hbox{\uwavefont\char58}%
    }%
  \ULon
}

\newcommand{\hlc}[2][yellow]{{%
                  \colorlet{foo}{#1}%
                  \sethlcolor{foo}\hl{#2}}%
}
\definecolor{quoteColor}{HTML}{FF8C00}
\newcommand\qt[1]{\hlc[quoteColor!30]{``#1''}}

\usepackage[normalem]{ulem}
\definecolor{linkColor}{HTML}{257E98}
\setuldepth{Berlin}

\newcommand\revised[1]{{#1}}

\newcommand{\lBox}[2]{%
  \fcolorbox{#1}{#1}{%
    \makebox[0.6cm][c]{%
      \textcolor{white}{\textbf{\textsf{\tiny \textsc{#2}}}}%
    }%
  }%
}

\newcommand{\lBoxSmall}[2]{%
  \fcolorbox{#1}{#1}{%
    \makebox[0.4cm][c]{%
      \textcolor{white}{\textbf{\textsf{\tiny \textsc{#2}}}}%
    }%
  }%
}

\definecolor{lgreen}{HTML}{3E8C75}
\definecolor{laqua}{HTML}{555A7B}
\definecolor{lblue}{HTML}{73A9B7}
\definecolor{lred}{HTML}{971E24}
\newcommand{\no}{\lBoxSmall{lred}{N}}
\newcommand{\wideNo}{\lBox{lred}{N}}
\newcommand{\yes}{\lBoxSmall{lgreen}{Y}}

\newcommand{\grammar}{\lBox{laqua}{DSL}}
\newcommand{\grammarAndGUI}{\lBox{laqua}{DSL+GUI}}
\newcommand{\gui}{\lBox{lblue}{GUI}}
\newcommand{\guiAndLibrary}{\lBox{lblue}{GUI+LIB}}

\newcommand{\code}{\lBox{lgreen}{LIB}}

\newcommand{\domainSpecific}{\lBox{lblue}{domain}}
\newcommand{\custom}{\lBox{laqua}{custom}}
\newcommand{\surveys}{\lBox{gray}{survey}}
\newcommand{\surveysAndCustom}{\lBox{laqua}{C \& S}}

\usepackage{xspace}
\usepackage[pagebackref,bookmarks]{hyperref}

\definecolor{steelblue}{RGB}{70, 130, 180}

\hypersetup{
    colorlinks=true, 
    linkcolor=steelblue, 
    urlcolor=steelblue,
    citecolor=steelblue
}

\let\oldhref\href

\newcommand{\ourhref}[2]{\oldhref{#1}{\textcolor{steelblue}{$\nearrow$ #2}}}

\begin{document}
\renewcommand{\figureautorefname}{Fig.}
\renewcommand{\sectionautorefname}{Sec.}
\renewcommand{\subsectionautorefname}{Sec.}


\vspace{-2em}
\firstsection{Introduction}
\label{sec:intro}

\maketitle

Experimental research is a mainstay method to infer causal relationships in visualization, HCI, and related areas~\cite{gergle_experimental_2014}. 
Experiments in visualization (often called user studies) range from perceptual studies \cite{cleveland_graphical_1984, szafir_modeling_2018}, to studies of visualization techniques \cite{ghoniem_readability_2005, kale_hypothetical_2018}, studies of interaction techniques \cite{boy_suggested_2016, kim_designing_2019}, and studies of full visualization systems \cite{nobre_evaluating_2020, lyi_learnable_2024}. Experimental approaches are also used outside of quantitative, controlled research, such as in eliciting expert feedback on systems~\cite{nowak_designing_2024} or designs~\cite{kieffer_hola_2016, he_design_2024, lisnic_visualization_2025}, or understanding insight formation~\cite{cutler_crowdsourced_2025, lee_how_2016}. 

While these studies historically have been conducted in the lab~\cite{cleveland_graphical_1984, zacks_bars_1999, dou_comparing_2010}, they are now predominantly run online through crowd work platforms~\cite{heer_crowdsourcing_2010, nobre_evaluating_2020, sukumar_designing_2018}, which support rapid participant recruiting and study execution. 
Most desktop-focused visualization tools are designed to be accessed through the browser, suggesting that web applications are an effective means to deliver study stimuli. These trends suggest that experimental research in the future will be predominantly conducted through browsers and asynchronously (as opposed to in a lab).

Despite the ubiquity of this form of inquiry, conducting these studies remains difficult for a variety of reasons. 
For instance, designing effective stimuli presents nuanced challenges (such as effective instrumentation of interactions) that are time-consuming to implement for experienced developers, let alone early-career researchers.
Similarly, the design of experiments offers a maze of complex decisions (such as choice of factors, randomization, sampling strategies, and partitioning of conditions) that, when navigated poorly, can invalidate entire experiments.
Compounding these design challenges is that making everything work as intended and ensuring that data is collected correctly can be tedious and error-prone.

To help with these challenges, researchers in visualization, HCI, psychology, and other fields draw on software frameworks to support their experiments.  
Whereas commercial tools, such as Qualtrics~\cite{qualtricsinternationalinc._qualtrics_2025}, excel at survey design, experimental design, and deployment, their closed-source and commercial nature hamper reproducibility, and they do not meet the more specialized needs of visualization experiments---such as sophisticated tracking of complex interactions. 
A variety of academic-led systems have been developed to help run studies; however, our discussions with stakeholders from the visualization community revealed that most are used only by individual research groups, use outdated technology, are difficult to maintain, and result in brittle deployments. Other academic systems are specific to certain visualization techniques (e.g., graphs~\cite{okoe_graphunit_2015}), or cover only a small slice of the experimental life cycle (such as factorial experiment design~\cite{musslick_sweetpea_2022}).

\pagebreak

To address these issues, we introduce reVISit 2, a software framework designed to support visualization researchers (and others) across the full life cycle of an experiment, including its design, debugging, and deployment. 
ReVISit is a proven system capable of supporting complex study designs at a publication-quality level. It has been used in various studies~\cite{lisnic_visualization_2025, cutler_crowdsourced_2025, cui_promises_2025, mcnutt_accessible_2025}. 
\revised{ReVISit 2 (referred to as simply reVISit throughout unless otherwise specified) builds on the foundation of reVISit 1~\cite{ding_revisit_2023}, but expands on our previous work in critical sociotechnical and technical dimensions.} For the former we have made a concerted effort to reach out to groups who may benefit from this tool via tutorials at universities and conferences, intentionally expanding and extending our documentation, and simplifying the startup cost for experiments by making common experiment components (such as VLAT~\cite{lee_vlat_2017}, Mini-VLAT~\cite{pandey_mini-vlat_2023}, BeauVis~\cite{he_beauvis_2023}, etc.) available as libraries. 

In addition to quality of life upgrades (such as improved UIs, form elements, data download, testing), on the technical side we make a variety of novel contributions to experimentation platforms. Central among these is an enriched domain-specific language (DSL) for specifying experiments, as shown in \autoref{fig:grammar-fig}. This language includes complex to implement features, such as Latin square participant distribution and randomized attention checks.
Complementing this expansion, we also provide high-level Python bindings, \emph{reVISitPY}, which offers Altair~\cite{vanderplas_altair_2018} style declarative specification and enables even more complex study designs (e.g., factorial designs) with little code.
A natural venue for use of this library is Jupyter notebooks, where we offer an experiment prototyping pipeline wherein developers cannot only design their experiment, but see and debug the experiment within the notebook, and even access collected data. 
This pipeline enables researchers to design and test their study, as well as prototype  analyses from within the notebook.

In contrast to other platforms, reVISit provides sophisticated tools for debugging and piloting, such as a study browser, participant view simulator, and data previews, as well as tools to manage data collection in ongoing studies.
We also make strides in simplifying the specification of visualization stimuli by making Vega visualizations first-class citizens, allowing us to provide automated provenance tracking across user interactions with Vega visualizations, which supports both low-level interaction analysis and fine-grained replay (\autoref{fig:texture-stim}).

We evaluate this collection of contributions via two strategies. 
First, we demonstrate its expressiveness through a collection of three study replications, where each study demonstrates different capabilities of reVISit (such as dynamic sequencing, capturing speech, and analyzing provenance).
Then we reflect on the design choices made in the system via three interviews with reVISit users and a close reading of the system lensed through Jakubovic \etals{}~\cite{jakubovic_technical_2023} Technical Dimensions of Programming Systems (TDPS). 
The combination of our technical design decisions and commitment to open source enables customized dissemination, which is useful during the review and reading process (reviewers and readers alike can explore studies easily and thereby build trust in results), as well as enables reproducibility, as data can be shared and studies can be forked easily.
Through this work, we seek to elevate the ease with which studies are conducted, improve the reproducibility and openness of studies within our community, and support the construction of advanced interactive studies.

\section{Related Work via Stages of a User Study}


\newcommand*\rot{\rotatebox{90}}

\newcommand{\myrot}[1]{\rot{\shortstack[l]{#1}}}
\begin{table*}[t]
    \small
    \centering
    \setlength{\tabcolsep}{0.9pt}
    {
        \begin{tabular}{lcccccccccccccccccll}
Name & \rot{Stimuli} & \myrot{Interaction\\Logging} & \rot{Libraries} & \myrot{Basic\\randomization} & \myrot{Dynamic\\randomization} & \rot{How to create} & \myrot{Study\\navigation} & \myrot{Records\\ response data} & \rot{Data preview} & \myrot{Video\\recording} & \myrot{Audio\\recording} & \rot{Precise Timing} & \rot{Eye tracking} & \rot{Other devices} & \myrot{Participant\\replay} & \rot{Open Source} & \rot{Active} && Specialty\\
\toprule
Google Forms~\cite{google_google_2025} \hspace{1mm} &  \surveys & \no & \no & \yes & \no & \gui & \yes & \yes & \yes & \no & \no & \no & \no & \yes & \no & \no & \yes & \\
psytoolkit \cite{stoet_psytoolkit_2017} &  \surveys & \no & \yes & \yes & \no & \gui & \no & \yes & \no & \no & \no & \yes & \no & \no & \no & \yes & \yes & \\
ETK~\cite{turton_etk_2017} &  \domainSpecific & \no & \no & \yes & \no & \code & \no & \no & \no & \no & \no & \no & \no & \no & \no & \yes & \no && Images\\
EvalViz~\cite{meuschke_evalviz_2019} &  \domainSpecific & \no & \no & \no & \no & \code & \no & \yes & \yes & \no & \no & \no & \no & \no & \no & \yes & \no && 3D Surface Visualizations\\
Flex-ER~\cite{lobo_flex-er_2020} &  \domainSpecific & \no & \no & \yes & \no & \grammar & \yes & \yes & \no & \yes & \no & \no & \no & \no & \no & \yes & \no && AR/VR\\
GraphUnit~\cite{okoe_graphunit_2015} &  \domainSpecific & \no & \no & \no & \no & \gui & \no & \yes & \no & \no & \no & \no & \no & \no & \no & \yes & \no && Network Visualizations\\
Evalbench~\cite{aigner_evalbench_2013} &  \custom & \yes & \no & \yes & \no & \grammar & \no & \yes & \no & \no & \no & \no & \no & \no & \no & \yes & \no & \\
Experimentr~\cite{harrison_experimentr_2019} &  \custom & \no & \yes & \yes & \no & \code & \no & \no & \no & \no & \no & \yes & \no & \no & \no & \yes & \no & \\
FROE~\cite{jansen_froe_2025} &  \custom & \no & \no & \yes & \no & \code & \no & \no & \no & \no & \no & \no & \no & \no & \no & \yes & \no & \\
VisUnit~\cite{jianu_visunit_2025} &  \custom & \yes & \yes & \yes & \no & \code & \yes & \yes & \no & \no & \no & \no & \no & \no & \no & \yes & \yes & \\
Touchstone1~\cite{mackay_touchstone_2007} & \custom	& \no & \no & \yes & \no &\gui & \no & \yes & \yes & \no & \no & \no & \no & \no & \no & \yes & \no & \\
PsychoPy~\cite{peirce_psychopy2_2019} &  \surveysAndCustom & \no & \no & \yes & \no & \guiAndLibrary & \yes & \yes & \no & \no & \no & \yes & \no & \no & \no & \yes & \yes && \\
jspsych~\cite{leeuw_jspsych_2023} &  \surveysAndCustom & \yes & \yes & \yes & \yes & \code & \no & \yes & \no & \yes & \yes & \yes & \yes & \yes & \no & \yes & \yes & \\
Qualtrics~\cite{qualtricsinternationalinc._qualtrics_2025} &  \surveysAndCustom & \yes & \yes & \yes & \no & \gui & \yes & \yes & \yes & \yes & \yes & \yes & \yes & \yes & \no & \no & \yes & \\
Sweetpea~\cite{musslick_sweetpea_2022} &  \wideNo & \no & \no & \yes & \no & \code & \no & \no & \no & \no & \no & \no & \no & \no & \no & \yes & \yes && Experimental Design\\
Touchstone2~\cite{eiselmayer_touchstone2_2019} &  \wideNo & \no & \no & \yes & \no & \grammarAndGUI & \no & \no & \no & \no & \no & \no & \no & \no & \no & \yes & \no && Experimental Design\\
ReVISit 1~\cite{ding_revisit_2023} &  \surveysAndCustom & \yes & \no & \no & \no & \grammar & \yes & \yes & \no & \no & \no & \no & \no & \no & \no & \yes & \yes & \\ \midrule
\textbf{\emph{ReVISit 2}} &  \surveysAndCustom & \yes & \yes & \yes & \yes & \grammar & \yes & \yes & \yes & \no & \yes & \no & \no & \no & \yes & \yes & \yes & \\
 \bottomrule
        \end{tabular}}
                \vspace{-1em}
        \begin{center}
        \surveysAndCustom \hspace{1mm}-- Custom and Survey, \code{} -- Library in General Purpose Language, 
        \grammar{} -- Domain Specific Language, \grammarAndGUI{} -- DSL and GUI   
        \end{center}
        \vspace{-2em}
    \caption{
    There are a wide variety of tools for online visualization-based experiments. These range from repurposed survey tools (\eg{} Google Forms, or similar tools not listed here such as SurveyMonkey~\cite{surveymonkey_surveymonkey_2025} 
    ) to domain-specific tools (\eg{} GraphUnit~\cite{okoe_graphunit_2015} for graphs or Flex-ER~\cite{lobo_flex-er_2020} for VR).
    }
    \label{tbl:study-tools}
    \vspace{-2.5em}
\end{table*}

Creating and conducting user studies is a complex process with iterated phases that each present challenges, as well as opportunities, for tool support. 
To focus such challenges, we divide the user study process into five stages, shown in \autoref{fig:teaser}, which we refer to as the experiment life cycle. 
In practice, this process is not linear but contains loops; for example, it is expected that after a pilot the experiment will be refined at the design stage to address issues that the pilot surfaced.
%

We detail each stage and situate ReVISit and prior work (particularly the tools in \autoref{tbl:study-tools}) within that stage.
\revised{We note that many tools are no longer maintained, limiting their practical utility.}
For a simple survey-based study, there are many easy to use open and commercial tools similar to Google Forms~\cite{google_google_2025}.
For custom experiments, jsPsych~\cite{leeuw_jspsych_2023}, reVISit, and Qualtrics~\cite{qualtricsinternationalinc._qualtrics_2025} present the most viable offerings---each catering to various audiences via differing approaches, as we discuss.


\subsection{Experiment Design}

In the experiment design phase, study designers decide \textit{what} participants see or do, and \textit{when / under which conditions} they see it. We separate this phase into two activities: stimulus and experiment design.
These designs are typically specified via a GUI, a library in a general-purpose programming language, or a DSL. 
These choices impact usability (e.g., GUI-based tools can be used by study designers with no programming skills) and expressivity (libraries and DSLs are potentially more expressive).
Notably, only commercial tools (Qualtrics, Google Forms, etc.) provide the means to specify studies via GUIs, which may be due to the development effort associated with GUIs. 

\parahead{Stimulus Design}
In most studies, a \textbf{stimulus} is presented to participants with the intent to elicit a response or behavior. In visualization research, stimuli typically take the form of images, text, video, audio (for sonification), objects (for physicalization), interactive applications, or combinations thereof. 
%
Digital stimuli are often designed to be viewed on a desktop screen, but other displays, such as AR/VR, large display environments, or mobile phones, are also used. In practice, tools that support non-desktop environments either are specifically focused on doing so (as in Flex-ER's AR/VR support) or are commercial tools (as in Qualtric's support for mobile).



Existing user study tools support a range of stimulus customization and features (\cf{} \textit{Stimuli} in \autoref{tbl:study-tools}). 
Several commercial tools are designed to conduct \textbf{surveys}, such as SurveyMonkey~\cite{surveymonkey_surveymonkey_2025},
and Google Forms\cite{google_google_2025}. 
Psytoolkit~\cite{stoet_psytoolkit_2017} is an open-source tool that also focuses on surveys. Although it is typically possible to include images and video as stimuli in a survey, they do not allow study designers to embed custom stimuli (such as web applications), and instead focus on the straightforward creation of form elements.  
Other tools focus on a \textbf{specific type of stimulus}, such as network visualizations (GraphUnit \cite{okoe_graphunit_2015}), images (ETK~\cite{turton_etk_2017}), 3D surfaces (EvalViz~\cite{meuschke_evalviz_2019}) or a specific modality, such as AR/VR (Flex-ER~\cite{lobo_flex-er_2020}). These tools offer customization options within their domain, but forgo arbitrary stimuli. 
However, their focus enables rapid stimulus creation within the domain.
%
Several tools enable study designers to integrate \textbf{custom stimuli}, typically in the form of bespoke UIs. Some tools enable combinations of custom stimuli with more structured form elements (\eg{} jspsych~\cite{leeuw_jspsych_2023}, Qualtrics~\cite{qualtricsinternationalinc._qualtrics_2025} and reVISit), and others require designers to write custom code for form elements (\eg{} experimentr~\cite{harrison_experimentr_2019}, FROE~\cite{jansen_froe_2025}, Touchstone1~\cite{mackay_touchstone_2007}). 

Related to stimulus choice is the decision of how to record responses and behaviors. Data is frequently collected via \textit{form elements} (\eg{} dropdowns, radio buttons, or text boxes), and is often encapsulated as common rating systems (\eg{} Likert scales~\cite{joshi_likert_2015}). 
Recordings of user behavior during the study are also common (as interaction logs or screen recordings), as well as more explicit user measurements media (such as through video or eye tracking). 
Qualtrics, jsPsych, Evalbench, and reVISit automatically log \textbf{browser events}, such as mouse movements or key presses.
Survey-centered and domain-specific tools make use of knowledge of their domain to automatically record data, whereas custom stimuli (typically being unconstrained HTML) require manual instrumentation. 
\revised{ReVISit includes automatically instrumented elements (forms and Vega programs) 
and means for instrumenting custom stimuli to support post-study replay and analysis.
}
\textbf{Audio recording} is rarely supported in the study platforms we surveyed; only jsPsych, Qualtrics, and reVISit have built-in audio recording capabilities, although recording could be implemented as part of a custom stimulus in many of the tools. Only reVISit offers automatic transcription. 

\parahead{Factors and Sequence Design}
Many studies test different conditions (independent variables), which in study design are commonly referred to as \textit{factors}. Typical factors in visualization studies are visual encodings, datasets, or tasks~\cite{jianu_visunit_2025}. 
For example, a study comparing node-link diagrams (NL) to adjacency matrices (AM) will vary the factors of visual encoding (NL, AM), datasets (\eg{} large vs. small, sparse vs. dense), and tasks (path finding vs. cluster identification).  
Two common designs use factors \textit{within subjects} (all participants see all values of a factor) and \textit{between subjects} (participants see a subset of factor values). \revised{Studies frequently combine \textit{within subjects} design for some factors and \textit{between subjects} design for others, to create \textit{mixed design} studies.}


Once each factor is decided on, designers need to consider the order and conditions in which stimuli are presented---the \textit{sequence}.
The simplest design is a \textit{fixed} sequence, wherein all participants see the same stimuli in the same order. 
Fixed sequencing may introduce confounders such as order effects (e.g., due to learning)\cite{gergle_experimental_2014}. 
To address this, it is common to (partially) \textit{randomize} the order in which stimuli appear, or in the case of between-subjects studies, randomly show different participants different stimuli. However, purely random distribution of stimuli may lead to limited coverage when many factors are considered with few participants. To improve balance \textit{Latin square} sequencing is often used, which ensures that stimuli will be seen equally frequently while controlling for order effects. 

Some studies implement more complex ordering that is not predetermined, and instead is dependent on the answers a participant gives during the study (e.g., to make the next question harder if the previous answer was correct). We call such sequences \textit{dynamic} sequences.  
For example, staircase designs~\cite{harrison_ranking_2014, yang_correlation_2019} involve iterated presentation of stimuli to find, for example, a perceptual threshold of some kind. 
Such dynamic sequencing is rarely supported, and often is implemented ad hoc (such as by embedding these designs into the stimuli themselves).

Some tools specialize in sequence design, while not providing support for other phases of the study life cycle. 
Touchstone2~\cite{eiselmayer_touchstone2_2019} features a GUI for creating and sharing sequence designs.  Complex sequence designs with multiple randomization strategies are creatable via GUI, and designs can be exported to configuration files that look similar to reVISit's.
Sweetpea~\cite{musslick_sweetpea_2022} is a Python library for sequence design that takes in a series of factors and produces experimental sequences, similar to reVISitPy.  Sweetpea includes a rich notion of constraints across factors that support more succinct expression of complex experimental designs than our current version of reVISitPy. 
VisUnit~\cite{jianu_visunit_2025} supports explicitly creating sequences from specified design factors (stimulus, dataset, tasks). 
Sweetpea, Touchstone2, and VisUnit do not support dynamic sequences, as their sequences cannot be adjusted based on user responses.
Like reVISit, jsPsych \cite{leeuw_jspsych_2023} supports dynamic sequencing via designer-defined functions that are called as trials get completed.

\subsection{Debug \& Pilot}

Once an initial study has been created and a design is decided, there is a phase in which study designers ensure that their stimulus, sequence, and data collection all work as intended. 
Debugging is commonly done by designers taking their own study (often many times) to test that everything behaves as intended. Efficient debugging requires being able to easily browse a study, without needing to take it from beginning to end.
GUI-based platforms typically provide the means to quickly navigate to a specific stimulus or condition, while library-based tools often do not provide such support and require developers either to know a URL or to take a study from beginning to end.

Piloting is done with participants who were not involved in the study (either via colleagues in so-called ``down-the-hall'' testing or through recruitment of preliminary participants online), and is done to identify problems, validate data collection, and collect preliminary data. Such preliminary data is often used to conduct a power analysis to estimate the number of participants required to find statistical significance of hypotheses \cite{cohen_statistical_2013}. 
Most tools do not provide advanced piloting support (\eg{} replays). ReVISit's support for this phase offers an important facet of our technical contribution.
Specifically, reVISit provides a study browser to navigate to different components with a participant view that shows a sequence that could be assigned, participant replays, and the ability for designers to take their own study to generate data.
\revised{Some tools (\eg{} jsPsych) can generate artificial data via simulation, which designers can use to identify experimental design problems and test analysis methods. ReVISit does not currently support simulation, however we believe that our DSL will make it straightforward to implement such a feature.}


\subsection{Data Collection} \label{data-collection}

The data collection phase begins with recruiting participants whose data is planned to be used in the final analysis. 
Participants can come from various sources, such as crowdwork platforms (\eg{} Amazon Mechanical Turk~\cite{amazoninc._amazon_2025} or Prolific \cite{palan_prolific.ac_2018}), volunteer-based or gamified platforms (\eg{} LabInTheWild~\cite{reinecke_labinthewild_2015}), and from networks such as mailing lists or social media. 
\revised{Although recruiting is largely orthogonal to study design, 
platforms such as Qualtrics offers (paid) access to participants, LabInTheWild provides basic study templates, and Prolific and Mechanical Turk have basic survey capabilities. }
During data collection experimenters must mind incoming data, which may involve rejecting fraudulent participants, identifying bugs, or examining initial data. 
Some tools have data previews (e.g., Qualtrics, Google Forms, and reVISit) that can be used for observability or simple analytics.

Once collected, this data needs to be stored somewhere.
Commercial tools typically provide data hosting as part of their service, sometimes for a fee.   
In contrast, non-commercial tools leave data hosting to the study designer.
For instance, reVISit primarily uses Google Firebase (a real-time-focused document database) for storage but does not provide a hosted solution. Instead, study designers must set up their own Firebase accounts---while the specifics differ, this is broadly typical.

\subsection{Analysis}

There are many sophisticated methods and tools to support the analysis phase of a study. As for participant recruiting, analysis tools are largely orthogonal to study frameworks.  
Statistical analysis is well supported by a multitude of environments and tools. 
For qualitative analysis, commercial tools (\eg{} MaxQDA \cite{verbisoftware_maxqda_2018}) are frequently used to help with the coding process. 
Various qualitative analysis tools have been created by the visualization community, such as VisTA~\cite{fan_vista_2020} or CoUX~\cite{soure_coux_2022}.
Other tools specialize in event sequence analysis~\cite{nobre_revisit_2021} (such as might be emitted by reVISit) or in analysis of eye tracking data~\cite{chen_gazealytics_2023}. 
\revised{While reVISit has some analysis capabilities, they are primarily designed to serve the debug/pilot and data collection stages---which is aligned with reVISit's design philosophy of providing functionality that is not already covered by high-quality open tools.
}







\subsection{Dissemination}


Faithful dissemination of study procedures, data analysis, and results is crucial for making results scrutinizable, reproducible, and ultimately building trust in the outcomes. 
A common approach to disseminate the details of a study is to include screenshots of the procedure or exports of a survey in a read-only format (e.g., from Qualtrics).
However, with web-based tools (see ``Open Source'' in \autoref{tbl:study-tools}), study designers can share both a link to the experiment as well as the code used to design it.
Study data is commonly shared via hosting platforms such as OSF or GitHub. ReVISit has a unique ability to share the data with the study, so that each participant's actions can be reviewed. For example, when including a screenshot of a response, authors can include a deep link to the stimulus page with a participant's actions, as in \autoref{fig:texture-stim}. 
\emph{We emphasize that reVISit is unique in its level of commitment to and extensive support of reproducibility and transparency}.

\subsection{Relationship to Previous reVISit Versions}




As the name would suggest, this work extends an earlier version of reVISit. 
A system that shares the name with reVISit focused on the analysis of user logs\cite{nobre_revisit_2021}. It did not provide study scaffolding, but instead explored topics such as event sequence analysis. It is the root of ideas for ``study rehydration'', i.e., the replay of a participant's analysis session provided in reViSit 2.
The reVISit study framework we report on here was started in 2022, with funding from the National Science Foundation. \revised{A VIS 2023 short paper \cite{ding_revisit_2023} describes the principles behind reVISit}: a DSL for defining experiments, components that contain stimuli, data collection that includes provenance tracking, and a process to compile everything into deployable web-based experiments. The first version recommended for public use, reVISit 1.0, was released in June 2024, followed by a 2.0 release in January 2025. 
We continued to expand on reVISit with crowdsourced think-aloud studies in a CHI25 paper~\cite{cutler_crowdsourced_2025}.
The key difference to prior versions is that reVISit now is a stable, well-documented, ready-to-use experimental platform, with early signs of community adoption (\autoref{sec:tdps-findings}). 
In addition, we make several technical contributions (Sec.~1).

\section{System Tour}

\revised{Next, we give a tour of ReVISit, highlighting notable features.}
%
In developing these features, we centered a design goal of making experiment design and deployment as frictionless as possible (for our target audience of technical scientists who might not be software engineers), while maintaining scientific sovereignty and without redoing what others already do well 
(\eg{} participant recruitment, data analysis).

For \textit{sovereignty}, ReVISit is deployed as a static web page (ensuring that there is no server to maintain). Deploying a static web page avoids vendor lock-in (users can just change the website as they see fit) and supports long-term dissemination stability (studies are static and do not change as reVISit changes unless the designer intentionally does so).  

For \textit{non-repetition}, we emphasize that ReVISit is not a database, a recruitment platform, a GUI experiment builder, or a single site analysis platform. Identifying that these are strengths of others, we design our system so that it takes advantage of those extant capacities, following recent guidance to ``lean on existing technological and social infrastructures''~\cite{akbaba_troubling_2023}. 
For instance, 
Prolific works well for study recruitment, and so we instead support the use of any recruitment platform;
statistical analysis tools in, e.g., R, are superior to anything we could provide, so we focus on compatible exports.  

Complementing these intents is a commitment to sociotechnical support, which we do via a mindfully maintained collection of artifacts (including \ourhref{https://revisit.dev/docs/introduction/}{tutorials, documentation, and examples}) as well as community efforts (such as a help forum and in-person tutorials).  

\subsection{The DSL}
\label{section-dsl}
The first step in setting up a reVISit experiment, after forking the base repo, is to start designing the experiment specification using our domain-specific language (DSL) (we give the grammar for this language in \autoref{fig:grammar-fig}). 
The root includes a list of named components that can be used in any sequence block, as well as a collection of component templates (\texttt{baseComponents}) used to partially define other components via inheritance.
This language
involves composing a collection of experimental ``blocks'' (in \texttt{sequence}).
Each block can contain stimuli (\texttt{components}) or nested blocks, as well as basic logic for controlling the order of components, and more fine-grained control, such as for inserting interruptions (such as for attention checks) and whether certain blocks should be skipped (such as due to wrong answers).



The reVISit DSL is a JSON-based DSL~\cite{mcnutt_no_2023}, in which experiments are specified through standalone JSON files.
These files are type checked through both a JSON Schema validation of the syntax as well as a secondary linter which identifies basic specification errors such as the presence of un-used components in the StudyConfig.

\newcommand{\opt}[1]{\textrm{#1}}
\newcommand{\oR}{\;|\;}
\newcommand{\anD}{,\;}
\newcommand{\funC}[2]{\emph{#1}_{\small\emph{#2}}}
\newcommand{\defined}{::=}
\newcommand{\secLabel}[1]{\emph{#1}}
\newcommand{\catLabelFree}[1]{\textrm{#1}}
\newcommand{\catLabel}[1]{\catLabelFree{#1}}
\newcommand{\partialComp}[1]{<#1>}
\newcommand{\varDef}[1]{\textbf{#1}}
\newcommand{\lbump}{\hspace{0.05in}}
\newcommand{\vbump}{\vspace{1mm}}

\definecolor{newElColor}{HTML}{A9E2C1}
\newcommand{\newEl}{\cellcolor{newElColor}}
\newcommand{\newHl}[1]{\colorbox{newElColor}{$\displaystyle #1$}}

\begin{figure}[t]
\small
\centering
$$
\begin{array}{l}

\lbump{}\varDef{StudyConfig} =\\
\left\{
    \begin{array}{rcl}
      sequence & \defined{} & Block\\
      components & \defined{} & \{ Name \rightarrow \partialComp{Component} \}\\
      \newEl{} importedLibraries & \newEl{} \defined{} & \newEl{} LibraryName[]\\
       \newEl{} baseComponents  & \newEl{} \defined{} & \newEl{} \{ Name \rightarrow \partialComp{Component} \}\\
      studyMetadata & \defined{} & ... 
    \end{array}
\right.\\

\noalign{\vskip 2mm}

\lbump{}\varDef{Block} =\\
\left\{
\begin{array}{rc@{}l}
    order & \defined{} & \opt{fixed} \oR{} \newHl{\opt{random}} \oR{} \newHl{\opt{latinSquare}} \oR{} \newHl{\opt{dynamic}_{\theta}}  \\
    components & \defined{} & (Name \oR{} \newHl{Block})[] \\
    \newEl{} numSamples & \newEl{} \defined{} & \newHl{\mathbb{N}^{+}}\\
    \newEl{} interruptions & \newEl{} \defined{} & \newHl{  (\opt{Deterministic}_{\theta} \oR{} \opt{Random}_{\theta})[]} \\
      \newEl{}  skip &\newEl{}  \defined{} & \newHl{  SkipCondition[]}\\
\end{array}
\right.\\

\noalign{\vskip 2mm}

\lbump{}\varDef{Component}=\\
\left\{
\begin{array}{rc@{}l}
\hspace{0.12in} compType  & \defined{}     & \opt{Markdown}_{\theta} \oR{} \opt{React}_{\theta} \oR{} \opt{Image}_{\theta} \oR{} \opt{Website}_{\theta} \oR{}    \opt{Form}_{\theta} \oR{} \newHl{\opt{Vega}_{\theta}} \oR{} ... \\
responses &\defined{} & Response[]
\end{array}
\right.\\
\noalign{\vskip 2mm}
\begin{array}{rc@{}l}

\varDef{Response} & \defined{}     & \opt{Numerical} \oR{} \opt{ShortText} \oR{} \opt{LongText} \oR{} \opt{Likert}  \oR{}  \opt{Dropdown} \oR{} \opt{Slider} \\ 
     & \oR{} & \opt{Radio} \oR{} \newHl{\opt{Video}} \oR{} \newHl{\opt{Checkbox}} \oR{} \opt{Reactive} \oR{} \newHl{\opt{Matrix}} \oR{} ... \\

\noalign{\vskip 2mm}

\newEl{}\varDef{SkipCondition} & \newEl{}\defined{}     & \newHl{\opt{BlockCondition}_{\theta} \oR{} \opt{RepeatedBlockCondition}_{\theta} \oR{} ...} \\
\end{array}\\
\noalign{\vskip 1mm}
\lbump{}\varDef{Name} = Symbol \anD{} \varDef{LibraryName} = Symbol, 
\textrm{\newHl{New Elements} in for ReVISit 2 }\\
\lbump{\theta\textrm{ denotes arguments, <X> is partial (or complete) definition \ala{} TypeScript}}
\end{array}
$$
\vspace{-2em}
    \caption{The reVISit grammar with configuration details elided.
    }
    \label{fig:grammar-fig}
    \vspace{-2em}
\end{figure}

Yet, specification through JSON is sometimes noted as being undesirable or messy syntax for DSLs~\cite{mcnutt_no_2023}.
Moreover, reVISit programs can be enormous---with some reaching tens of thousands LOC due to repetition of structures to combine multiple factors.
To address these issues, we developed a Python wrapper for the reVISit DSL called \ourhref{https://pypi.org/project/revisitpy/}{reVISitPy}.
These Altair~\cite{vanderplas_altair_2018} style bindings allow study designers to make use of the full expressivity of general-purpose languages---allowing for variables, complicated looping logic, and so on.
In the interest of keeping the library's syntax familiar to visualization developers (who would likely also be study designers), we intentionally mimicked Altair's use of structured-like method chaining. 


Echoing how tools like Altair are often used in the context of Jupyter notebooks, reVISitPy is designed to work well within notebooks. 
\revised{To this end, reVISitPy supports in-notebook previews of the experiment and the collected data. Test data can be retrieved from the preview, so that data wrangling and analysis can also be prototyped in the same notebook---see the appendix or this \ourhref{https://revisit.dev/docs/revisitpy/examples/example_jnd_study/}{example}.}
This supports rapid workflows wherein the experiment designer composes a reVISitPy program, views the effects of their design, and makes iterative adjustments. 

Some extremely custom designs or those implementing high-level constraints (\ala{} Sweetpea~\cite{musslick_sweetpea_2022}) are more straightforward to express via direct specification of the JSON DSL. However, the Python bindings aim to simplify the process of making small prototypes and simplify experimental design more generally by keeping it centered in a single computational notebook.
For example, during a team retreat, we prototyped our JND replication (\autoref{sec:jndstudy}), from dataset generation, to stimulus generation (via Altair), to experiment specification, testing and piloting, and preliminary analysis, all from within a single notebook.



\subsection{Stimuli}
To a study participant, the most evident part of an experiment is the stimuli that they interact with. 
ReVISit experiments can include a variety of types of stimuli, including form elements (numerical inputs, sliders, etc.), markdown files (such as for participant instructions, consent forms, and so on), images, and videos. 
Naturally, any list of prebuilt components will be incomplete, and so we support custom components by allowing users to supply generic web-components as well as React components which can be smoothly integrated into reVISit's full data and provenance tracking capabilities by use of the trrack library~\cite{cutler_trrack_2020}.

However, merely exposing an endlessly customizable component takes the focus off of visualization. 
In experiments---and visualization practice more generally---a common way to create visualizations is through the use of DSLs, such as Vega~\cite{satyanarayan_vega-lite_2017}. These DSLs simplify the specification of often repetitive structures (such as data management or scaling code). 
Echoing this approach, we include Vega programs as first-class stimuli, allowing them to be specified directly in the reVISit DSL or imported from a separate static file. 
We automatically instrument these programs with provenance tracking, such that the state of the Vega signals is recorded as users interact with the programs.
Fine-grained state tracking supports similarly fine-grained participant replay (as we discuss below), such as being able to view specific hover states and mouse moves. 
Further, consistent with other custom components, we offer a custom Vega signal callback that can be called to set the answer inside of reVISit's reactive responses while using Vega, which allows a user to interact with the visualization and click on elements to set the answer that will be recorded by reVISit.

\begin{figure}
    \centering
    \includegraphics[width=\linewidth]{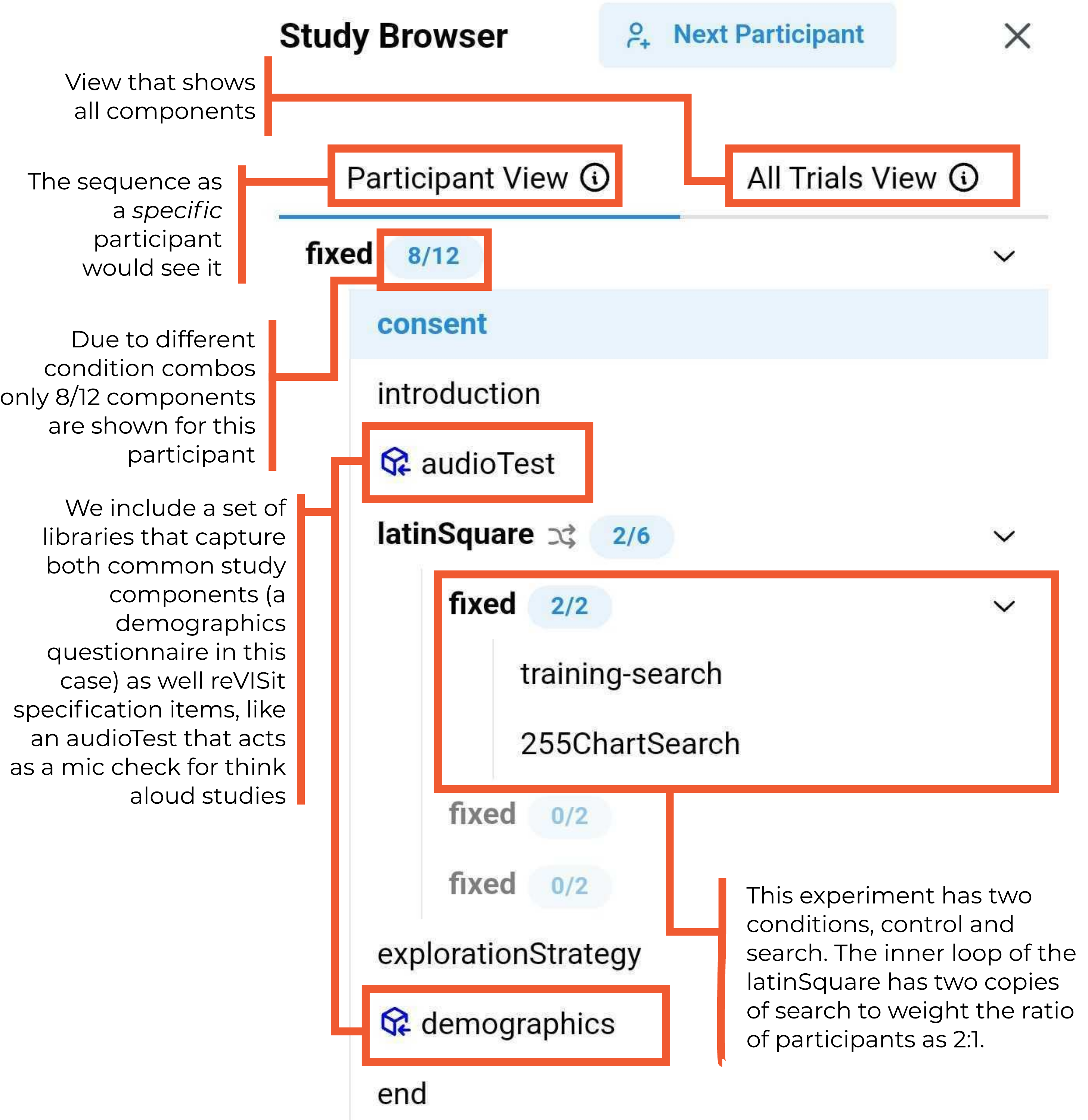}
    \vspace{-2em}
    \caption{The study summary for our search study replication. 
    Each participant flows through the experiment, first seeing consent, then introduction, and then is sorted into one of three conditions (the fixed subsections of the Latin square). This summary can be seen \ourhref{https://revisit.dev/replication-studies/255chart-study}{here}.}
    \label{fig:study-browser}
    \vspace{-2em}
\end{figure}

Many experiments use standardized surveys or other common stimuli as part of their design---\eg{} VLAT. 
To support this usage, reVISit provides a collection of libraries that support various common tasks (as in \autoref{fig:study-browser}), including demographics questionnaires, color vision deficiency tests, or visual literacy tests~\cite{lee_vlat_2017, pandey_mini-vlat_2023}. 

\subsection{Sequence}
The next aspect apparent to a participant is the order in which components appear.  Each \textit{block} in our DSL has a defined \textit{order} in which its child components are shown. \revised{ReVISit includes affordances for specifying the order and relationship between components that support rich customization and experimental designs (\eg{} between-subjects, within-subjects, and mixed designs)}. 

\texttt{Fixed} order shows components in the order in which they are listed, whereas \texttt{random} shows them in a random order (per participant). 
Yet, in studies with limited participants and many conditions, random ordering does not guarantee sufficient coverage of the study cases. 
\texttt{Latin square} orderings are commonly used to provide such guarantees~\cite{gergle_experimental_2014}, being particularly useful as a way to mitigate order effects.

While these strategies cover many different designs, they do not capture all possible sequences. 
Some studies rely on the answers to previous questions to determine the next stimulus that will be shown to a participant. These cannot be specified in our DSL and require custom logic to implement. 
To enable more nuanced types of designs we support dynamic ordering, in which a study designer provides a bespoke JS function that is called repeatedly as the participant progresses through a block to determine their next task.
For example, consider a study design where participants see different stimuli based on the accuracy of their earlier answers---such as how US-based Graduate Record Examinations (GRE) adaptively alters the difficulty of topic sections based on the success rate of previous sections (see \autoref{sec:jndstudy} for an example). 
Dynamic ordering could be accomplished within a stimulus; however, using dynamic functions enables various useful reVISit features (\eg{} logging, participant replays, or stimuli navigation). 


In addition to linear progression through an experiment, some studies may require periodically inserted components separate from the experiment logic (such as attention checks) or non-linear jumps (such as ejecting a participant if they fail a training).
Each of these tasks are supported by blocks through their interruption and skip logic, respectively. 
Although these functionalities could be orchestrated through a collection of dynamic checks and custom components, we elevate these to language-level features to highlight their importance in study design. 


We argue that reVISit can model more diverse study designs than alternatives, such as Qualtrics~\cite{qualtricsinternationalinc._qualtrics_2025} or VisUnit~\cite{jianu_visunit_2025}. 
For example, VisUnit lists staircases as a design form that it cannot model. 
ReVISit can model such designs, as in our staircase-based replication in \autoref{sec:jndstudy}, which is enabled by our robust native and dynamic sequencing.


\subsection{Study Browser} 
\label{sec:study-browser}

Understanding the architecture of a study, such as which blocks are contained within which other blocks or which stimuli will be seen by what fraction of the population, as well as navigating to specific stimuli, are difficult challenges in the debug \& pilot phase of an experiment.
ReVISit addresses these through a \textit{study browser}, shown in \autoref{fig:study-browser}, which organizes experiments into a single summative view describing the mechanical architecture of the study. 
The table of contents-like structure is situated on the right hand side of the application while in ``admin'' mode, juxtaposed to the rest of the experiment (as in \autoref{fig:texture-stim}-A).
Clicking on a stimulus instantly navigates to that stage of the experiment. For instance, clicking on ``255ChartSearch'' in \autoref{fig:study-browser} brings up the specific search stimuli of interest (\cf{} \autoref{fig:search-distribution}), thereby speeding up debugging. 

The study browser has two views: a ``Participant View'' (active in \autoref{fig:study-browser}), which shows the sequence just as a particular participant would see it, including the order and the selected subset of trials. Clicking on ``Next Participant'' rebuilds the sequence for another participant, thereby enabling experiment designers to check that all sequencing is specified correctly. The ``All Trials View'' makes all components immediately accessible, independent of whether they appear in the sequence of a particular participants, enabling designers to quickly navigate to each component.  
Finally, the study browser surfaces response data (\eg{} right or wrong) when replaying a participant run.




\subsection{Data}
\revised{After the study is in place and launched, data collection commences. ReVISit  supports several \textit{storage engines}, including browser-storage, Google Firebase, and Supabase; custom storage engines can be implemented by interested users. We plan to expand our existing storage engines to support other common data hosting services in the future.}
In addition,
ReVISit includes a variety of different affordances to tend to this data across the experiment life cycle. 
\revised{Although we refrain from re-implementing mature analysis tools, some analysis steps are better situated within the system rather than as an external analysis loop. After the experiment, data can be exported (\eg{} as JSON or CSV) and used in custom analysis workflows.
}

The first analysis feature is \textbf{participant replay}, which allows study designers to watch individual participant trials. ReVISit's rich notion of provenance allows for straightforward rehydration of study stimuli (such as the Vega stimuli described above), such that individual actions like keystrokes in a form can be observed, as can be seen in \autoref{fig:texture-stim}-A. The timeline and rich event logs at the bottom enable analysts to investigate each step a participant took, supporting analysis of the data---for instance in our search replication (\autoref{sec:search-study})
---as well as debugging of pilot studies.
These interactions are reified as our \textbf{timeline view} (\autoref{fig:texture-stim}B), which shows a timeline of all of the tasks a participant took across an experiment, and also shows responses on tooltip. \revised{
While we record mouse movement, variation in monitor sizes and aspect ratios make visualization of mouse movements difficult.
}



A related useful form of data for the prototyping phase is the \textbf{audio solicited through think-aloud studies}. Think-aloud can be used for usability evaluation and insight elicitation~\cite{cutler_crowdsourced_2025}, such as in our search replication (\autoref{sec:search-study}). A novel usage would be to allow pilots to verbally describe their thought processes as they are doing it---rather than requiring them to recollect after the study or take notes. 

Lastly, we provide a collection of simple analytic views that allow for quick sanity checking about statistics in the experiment. 
These include a minimalist tabular view for spot checking the data and simple analytics views that summarize participant performance by trial.
The intent of these views is analogous to how Google Forms offers a collection of basic summative charts about the collected data as a way to check the distribution of things rapidly. 

\subsection{Post Study}
After a reVISit study is complete, the work it needs to do is not entirely finished. Concerns related to ensuring the long-term accessibility, transparency, and replicability of the experiment are central to reVISit's design and so we offer a collection of features to support these tasks. 


The first focus is on the explainability of the experiment, which reVISit supports by helping reviewers and other interested parties understand the study and its results. 
We optionally allow non-verified visitors to navigate the experiment without having to fully take the study (as a curious reviewer might wish to do)---such as at \href{https://revisit.dev/replication-studies/HeatmapJND-study}{revisit.dev/replication-studies/HeatmapJND-study}.
We also provide an option to deactivate data collection so as to avoid collecting data from reviewers or other unwanted sources.
Study admins can also give access to the same data downloads, as well as timeline and replay visualizations, that study creators had. Similarly, reVISit enables authors to deep-link into individual participants' trials, as we do for all reVISit figures (such as \autoref{fig:texture-stim}), enabling readers to scrutinize the context. 

The second focus is long-term reproducibility. As noted above, the intended workflow for reVISit is for study creators to fork the GitHub repository, thereby creating a snapshot of the current version of the reVISit code, which will remain stable despite updates to the main repository.
Additionally, having all of the code required to run the study in the same repository, including stimulus, the reVISit tool, and the configuration for experiment design, is designed to make future replications or modifications to the study simple. 
Although there are vulnerabilities to this architecture---such as library dependencies catastrophes~\cite{abdalkareem_impact_2020} or changes in browser functionality---this approach offers substantially more straightforward access to experiment stimuli and architecture than comparable closed-source alternatives.  
We intend to continue to enrich the suite of features offered in this area, such as automatically creating archival screenshots and videos of the experiment that will be fully resilient to changes in browser technologies. 







\section {Field Test: Replications}

To demonstrate the utility of reVISit on real-world experiments, we replicated three extant studies~\cite{harrison_ranking_2014, he_design_2024, feng_effects_2018}. 
We selected these studies to demonstrate different aspects of reVISit---for instance, the first study demonstrates our dynamic sequence capability. 
In addition, we add new variations to each replication---such as by testing additional conditions or leveraging additional data collection modalities---so as to not merely replicate but to expand the previous studies. In total, we recruited 460 participants for the studies, of which 440 were recruited via crowdsourcing, and 20 were visualization design experts recruited via social media.  All studies were pre-registered on OSF, and are available at \href{https://revisit.dev/replication-studies/}{revisit.dev/replication-studies}. In this section, we sketch the studies and key results, while focusing on lessons learned as each study was executed in reVISit. More details are available in the Appendix.


\newcommand{\jndstudytitle}{Replication of Ranking Visualizations of Correlation}
\subsection{\jndstudytitle}
\label{sec:jndstudy}
\begin{figure}[!t]
    \centering
    \includegraphics[width=\linewidth]{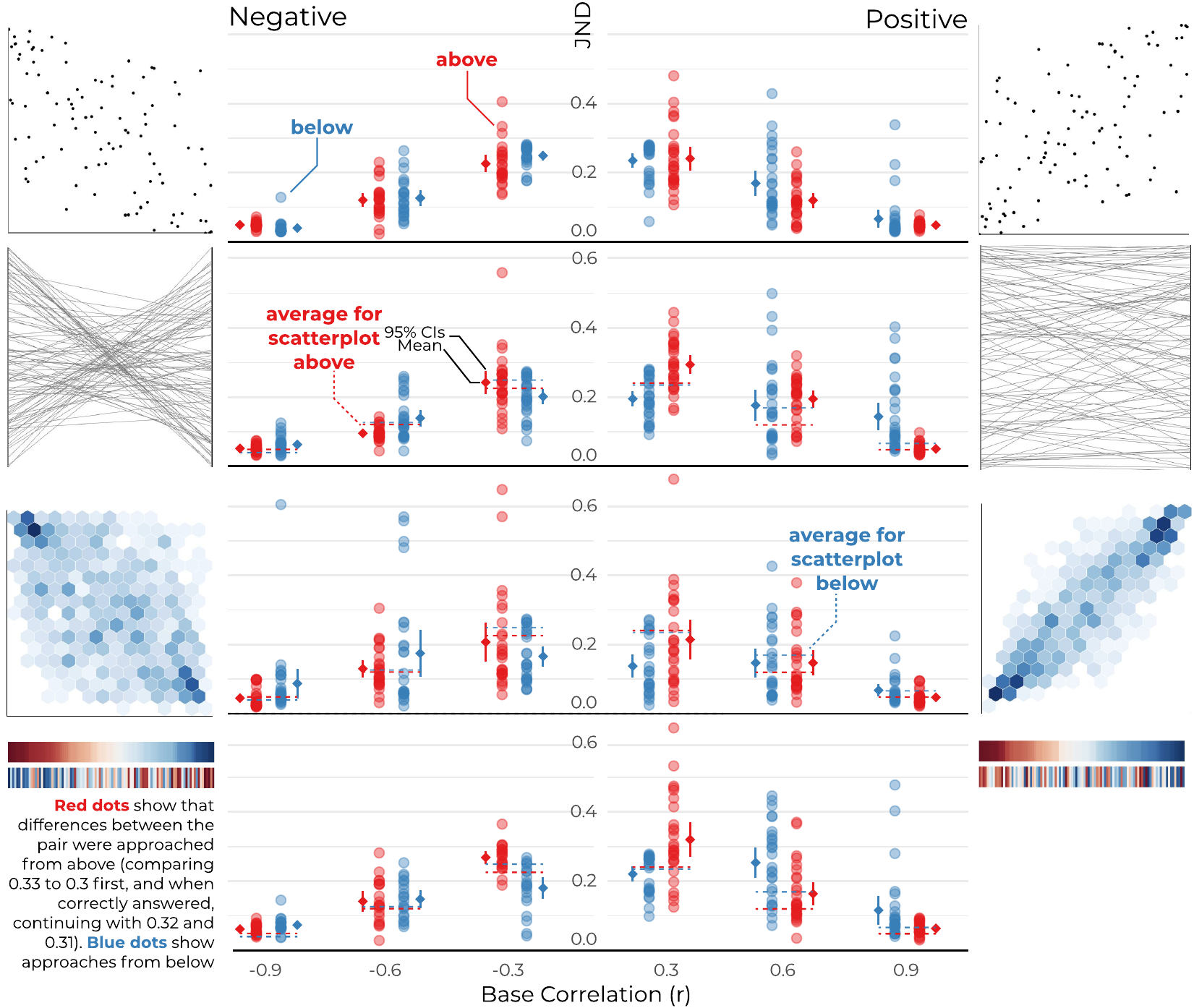}
    \vspace{-6mm}
    \caption{   
    Distribution of JNDs across eight conditions---four types of visualizations and positive and negative correlations for each. Scatterplots are best at showing differences in correlation. Hexbin plots show individual differences (see outliers), but are best at showing differences for low r values. Lower JNDs indicate easier detection of correlation differences. JNDs are generally lower for high correlation values. 
    }
    \label{fig:jnd-distro}
    \vspace{-2em}
\end{figure}

We replicated Harrison et al.'s~\cite{harrison_ranking_2014} study on just noticeable differences (JND) of correlation values in varied visualization techniques.
Harrison \etal{} found that different chart forms vary in how precisely participants can distinguish two similarly correlated plots of two-dimensional data (scatterplots work best), and that the JND is smaller when the correlation values $r$ are higher. 
We run the study with two already examined techniques (scatterplots and parallel coordinate plots) as well as two new ones (hexbin plots and sorted heatmaps). 
We test each condition with different correlation values,
(see \autoref{fig:jnd-distro}). 
As in the original study, we use 100 points in each condition, except for hexbins, which are typically used for larger datasets, where we use 1000.

We selected this experiment because (a) it uses a challenging experimental staircase design that is not typically supported by experimental frameworks, and (b) we were interested in expanding the set of tested visualization techniques.
We used reVISit's dynamic sequencing function to implement a staircase design, adjusting follow-up pairs of correlations based on previous responses. We also terminated trials early based on statistical tests of performance. We added an attention check involving an obvious pair of correlation values ($r=0.01, 1$) placed so that we could detect thoughtless clicks. 
We recruited 30 participants using Prolific for each condition, totaling 240 participants.

\parahead{Results}
Our experiment replicates Harrison \etals{} findings for scatterplots and PCP plots (see \autoref{fig:jnd-distro}). Our results show even narrower confidence intervals, which may be attributable to the improved attention check or to the other advantages of our experiment, such as more robust data quality control and a more modern interface.
We find that hexbin plots are suitable to visualize correlations for large datasets, and seem to even outperform scatterplots for low correlation values.
However, several poorly performing participants produced outliers, which indicates that hexbins might not be universally understood. 
Heatmaps perform the worst of all stimuli, although not disproportionately,  suggesting they may be useful for space-restricted mediums. 


\parahead{Lessons Learned}
For our study design, we opted to run the conditions as separate studies on Prolific, rather than assigning conditions via reVISit within a single study. An alternative Latin square design within one study can become unbalanced based on returned studies on Prolific, which we plan to address in the future with a plug-in for Prolific, so that a Latin square entry would be made available again when a participant returns a study. Separating the conditions into different studies granted us more control over participant numbers per condition, but came with tradeoffs as we had to duplicate study configurations across the conditions. Although we could have leveraged reVISitPy to generate multiple study configurations and reduce overhead, more built-in support for between-study design may be useful. 


We implemented the staircase procedure using the dynamic functionality in reVISit, rather than embedding logic directly in the stimulus, as the original study did~\cite{harrison_experimentr_2019}. 
ReVISit’s dynamic functions allowed us to develop, debug, and pilot studies consistently across all conditions and maintain a clean and centralized implementation of the staircase logic. 

\begin{figure}
    \centering
    \includegraphics[width=\linewidth]{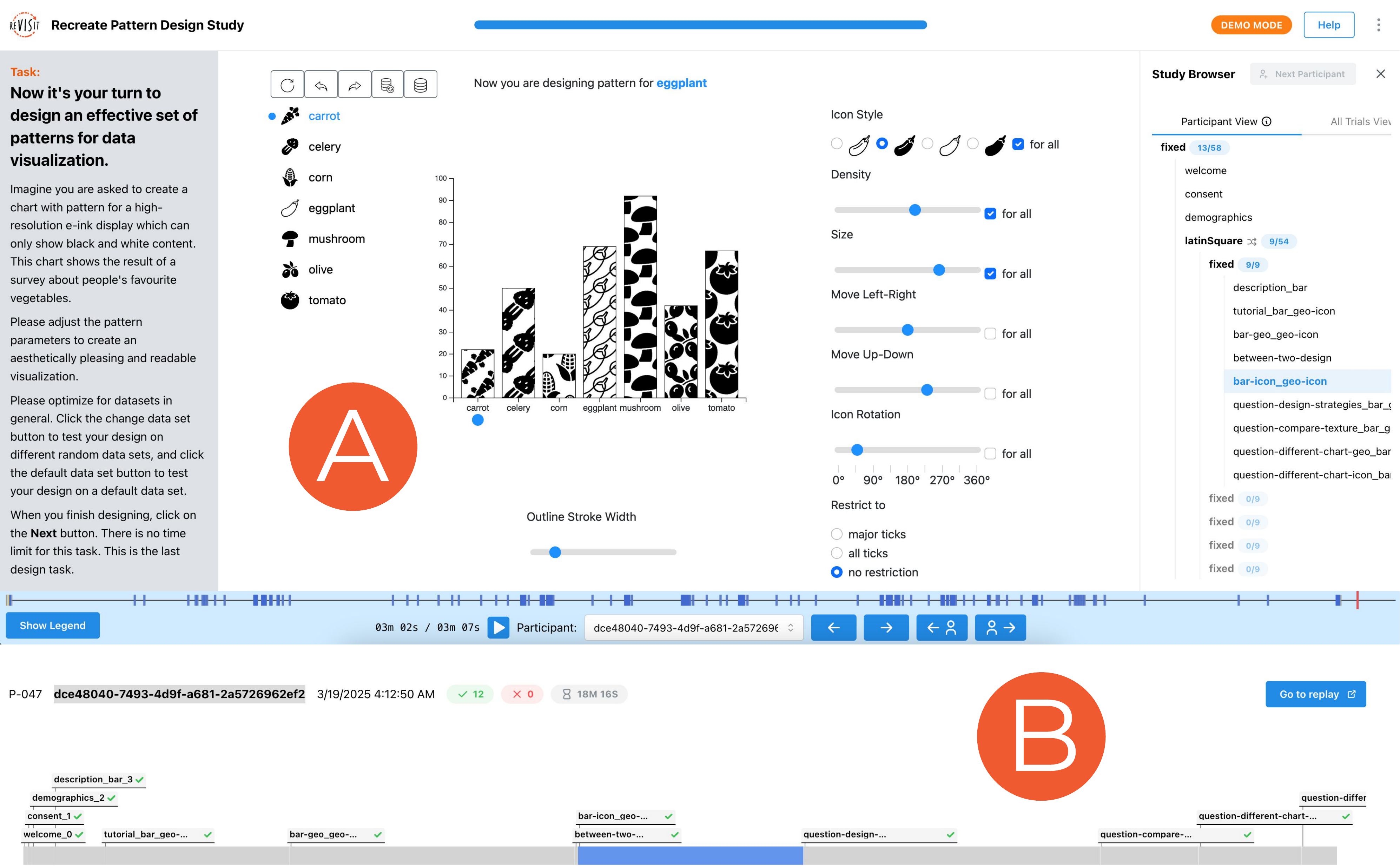}
    \vspace{-4mm}
    \caption{A participant's result from our texture study shown in analysis mode. (A) Participant replay shows events and navigation options at the bottom. (B) The participant timeline gives summary information and serves as an entry point. [\ourhref{https://revisit.dev/replication-studies/pattern-design-study/V3ZPZStsZmxBWDNta2p4MyttU1dJZz09?participantId=dce48040-7493-4d9f-a681-2a5726962ef2}{View Result}] }
   
    \label{fig:texture-stim}
    \vspace{-2em}
\end{figure}

\newcommand{\patternstudytitle}{Replication of Pattern Design Study}
\subsection{\patternstudytitle}
\label{sec:pattern-study}

Next, we replicated He et al.'s study \cite{he_design_2024} on designing black-and-white patterns for visualization. They asked visualization experts to create and then rate designs for different patterns (geometric and iconic) and chart types (bar, pie, or map). The authors then coded participants' responses and categorized the goals into those related to readability and aesthetics, and summarized corresponding design strategies.

We chose to replicate this study to show that reVISit 
(a) supports a mixed study design (between-subject design on chart type and within-subject design on pattern order) and (b) that existing tools can be integrated into the reVISit framework. In contrast to the previous study, we fully instrumented the stimulus so we could analyze design strategies and reason about the number of interactions associated with good or bad designs. We judged each design internally (via the BeauVis scale~\cite{he_beauvis_2023}), in place of a follow-up crowdsourced study as the original did.
We recruited 20 participants (design experts) from social media.

\begin{figure}[t]
    \centering
    \includegraphics[width=1\linewidth]{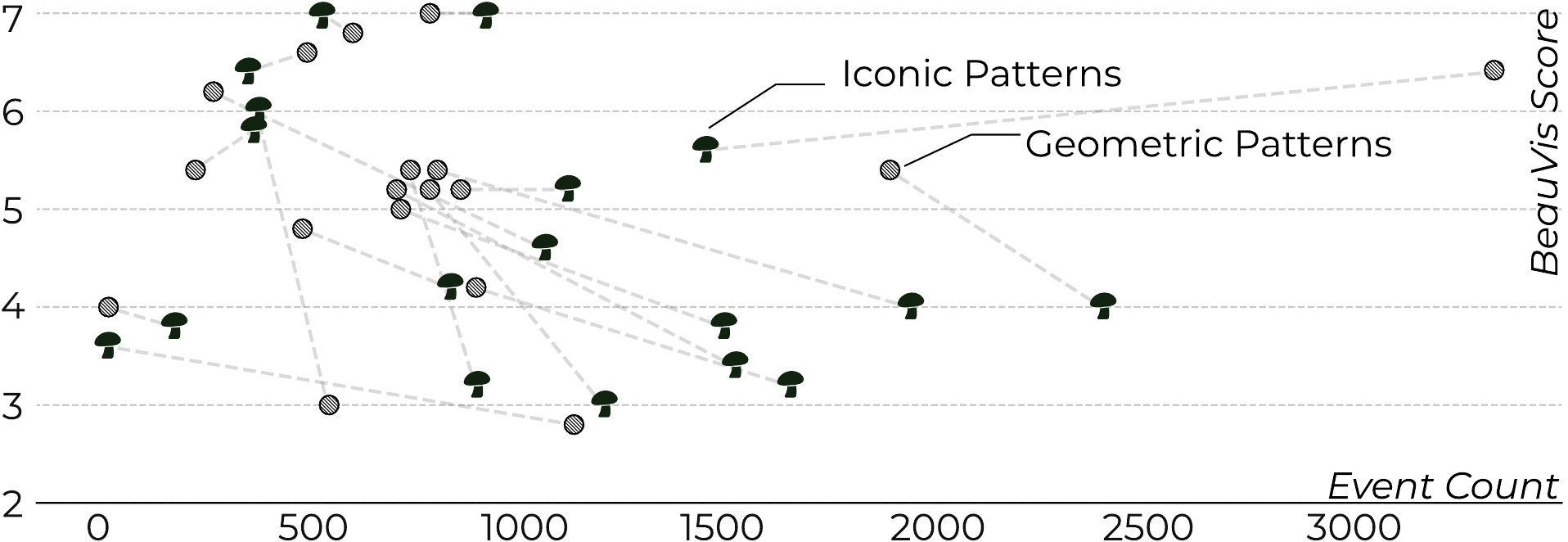}
    \caption{
    Relationship between interaction events and BeauVis~\cite{he_beauvis_2023} scores in our pattern study. Each pair of points shows a single participant.
    }
    \label{fig:pattern-beauvis-events}
    \vspace{-2em}
\end{figure}

\parahead{Results}
Our results replicated He \etals{} in finding similar design goals and design strategies.
Participants' design goals focused on distinguishability (15\texttimes{}), visual clarity (5\texttimes{}), semantic association (5 \texttimes{}), visual pleasure (7\texttimes{}), and visual balance (3\texttimes{})---see \autoref{fig:pattern-beauvis-events} for a distribution of scores. 
Two new strategies emerged: 
one participant attempted to create new shapes by playing with the basic geometric shapes to create new shapes,
and another designer experimented with dot density to create different shades of gray, \ala{} halftoning.

Through the reVISit replay interface, we discovered common strategies that designers use. 
A typical workflow involved an initial exploration of different default patterns, followed by iterative refinement of individual patterns, and finally testing and refining the design with different datasets (e.g., \ourhref{https://revisit.dev/replication-studies/pattern-design-study/V3ZPZStsZmxBWDNta2p4MyttU1dJZz09?participantId=8fb95e0a-c455-42e5-8048-b28dc7b56230}{as this participant did}). 
For iconic patterns, a commonly observed behavior was testing different icon styles (like \ourhref{https://revisit.dev/replication-studies/pattern-design-study/V3ZPZStsZmxBWDNta2p4MyttU1dJZz09?participantId=dce48040-7493-4d9f-a681-2a5726962ef2}{this participant did}). 
We noticed these design strategies only by using participant replay, highlighting the replay's value in uncovering subtle yet impactful design behaviors.

\parahead{Lessons Learned}
In our study configuration, we used a Latin square to order components. 
However, we later realized that this approach is not well-suited for studies recruiting participants via social media, as we did. A reVISit Latin square is calculated based on the participants who start the study, rather than those who finish it---each initiated study takes an entry from the Latin square. However, many social media users click on the study link but do not complete it, leading to an imbalance between conditions. 
Although study designers can manually reject incomplete entries to rebalance, we plan on adding a timeout mechanism that automatically rejects incomplete entries and returns their entries to the Latin square. 
We recommend that studies with small sample sizes or recruiting via social media use random orderings.

Instrumenting an existing interactive visualization tool with Trrack~\cite{cutler_trrack_2020} to enable full rehydration required considerable effort.
To support replay, we needed to capture every user interaction and reconstruct the visual state of the application from the captured provenance data, which required building a stable history management system. Most existing visualization libraries, however, do not support such functionality \cite{zhao_libra_2025}. 
To lower the burden when using a legacy system, a combination of screen capture and logging could reduce the technical burden while achieving most of the functionality of full rehydration.


\subsection{Replication of Search in Visualization Study}
\label{sec:search-study}

Finally, we replicated Feng \etals{} study \cite{feng_effects_2018} on search functionality in interactive visualizations.
The study involved three visualization types, each tested under two conditions: (a) with search functionality and (b) without it. 
Participants were asked to analyze the visualization and document their findings. 
The study tracked their interactions, including total time spent on each visual element, whether they used the search function (if available), and whether their interactions involved searched items.
The study found that the presence of text-based search influenced participants' information-seeking behavior. 
Specifically, search functionality encouraged users to actively seek individual data points and spend more time examining details within the data.

We chose to replicate this study because 
(a) it utilizes interaction logging and 
(b) it elicited information about participants' process during the task, but was constrained to text-based input after the exploration task. We alternatively elicit think-aloud data from participants during the task in our replication.
We follow the original study design by replicating two of the three visualizations with minor modifications%
---a bubble chart, and NYT's ``255 charts''~\cite{jeremyashkenas_how_2014}. 
The original study was implemented using vanilla JS and D3. 
We copied over the original code for 255 charts into reVISit using a website component, whereas we re-implemented the Bubble Chart in React. 
\revised{We recruited 99 and 93 participants for the Bubble and 255 charts conditions respectively after applying the original study's minimum interaction exclusion conditions.}

\begin{figure}
    \centering
    \includegraphics[width=\linewidth]{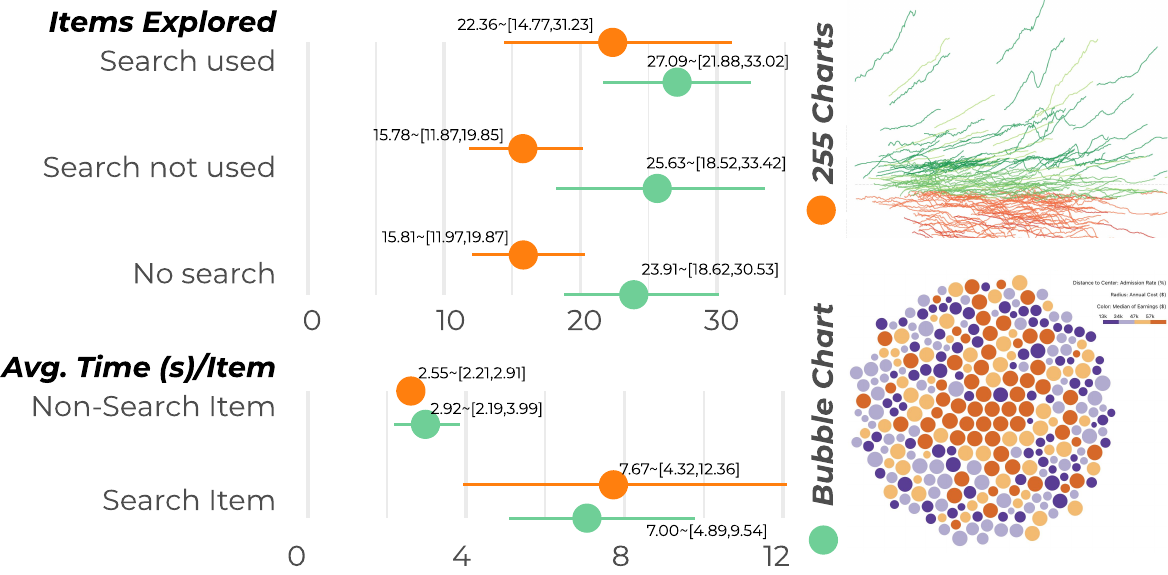}
    \caption{Replication results from the search study align with prior findings (not shown), suggesting that while participants tend to interact with similar numbers of items, when participants engage with items highlighted during search, they tend to interact with them for longer.}
    \label{fig:search-distribution}
    \vspace{-2em}
\end{figure}

\parahead{Results}
Our experiment replicates Feng \etals{}~\cite{feng_effects_2018} findings, in that participants in both conditions explored similar numbers of items (\autoref{fig:search-distribution}). 
However, the main effect of items highlighted in search being engaged with for longer also holds across both conditions.
For example, in the Bubble visualization, items highlighted by search were statistically significantly  viewed longer 
($M(ean) = 7.00$s) than those not highlighted ($M = 2.92$s) via mouse click. 
Similarly, in the 255 charts visualization, search-highlighted items were also statistically significantly viewed longer 
($M = 7.67$s) than non-search items ($M = 2.55$s) among search users. 

\parahead{Lessons Learned}
ReVISit's provenance tracking allowed us to measure where/how-long participants were interacting, but also allowed us to recreate their exploration session for deeper qualitative insights. 
Combined with the audio-enabled think-aloud protocol, this led to several new insights. 
For example, we can observe how participants ``orient'' themselves in the visualization: \ourhref{https://revisit.dev/replication-studies/bubblechart-study/LzE2MTl4ZVRMTk5nSFlNYmd1ZDhjZz09?participantId=d90ad824-3bc1-4b7d-8740-38827c7f86ab}{``So, this one is relatively small, the circle. And it looks like it's an orange circle, which would appear that it's on the higher end of earnings for people that attend this school''}.
It also confirmed a hypothesis from the original paper: that people would often use search to examine data that was personally relevant to them (\ourhref{https://revisit.dev/replication-studies/255chart-study/LzE2MTl4ZVRMTk5nSFlNYmd1ZDhjZz09?participantId=3a12c79b-75ce-4c1a-82ed-8b95f8523ae7}{in this case, particular industries}).

\section{Reflection}

Next we reflect on the design of our system. To do so we interviewed users of reVISit and did a heuristic analysis of the system.


\parahead{Interviews} We conducted a semi-structured interview study with researchers who have used reVISit in a completed study.
Although reVISit has been used many times~\cite{cutler_crowdsourced_2025, lisnic_visualization_2025, cui_promises_2025, mcnutt_accessible_2025}, there is often an overlap between the maintainer of the system and the researchers running the study, in part because of our persistent community-centered outreach strategy. All participants were current PhD students who have run multiple user studies as part of their PhD, have run at least one study with reVISit, and are not part of the reVISit development team. 
We refer to participants as PX and \qt{quote them}. 

\parahead{Technical Dimensions (TDPS)} We conducted a close reading of reVISit via Jakubovic \etals{} ``Technical Dimensions of Programming Systems (TDPS)''~\cite{jakubovic_technical_2023}. TDPS is a collection of dimensions (\ala{} Cognitive Dimensions of Notation~\cite{green_cognitive_1989}) by which to understand systems across 7 clusters of dimensions. 
The first author characterized how reVISit addresses each dimension, which was subsequently reviewed by two other authors. We use McNutt \etals{}~\cite{mcnutt_mixing_2025}  cluster descriptions. We elide \textbf{\emph{Conceptual}} (as it is broadly covered in our discussion of the DSL)  and \textbf{\emph{Complexity}} (which largely overlaps with notation).


\subsection{Findings}
\label{sec:tdps-findings}

\parahead{Interaction}  \textit{Which loops in the system are overlapping and how far apart are the corresponding gulfs of evaluation?} A common problem in developing user study prototypes is wide feedback loops during the development and testing of a study. Our study browser, as well as our development environment, is primarily aimed at tightening these feedback loops. After forking our repo, designers are met with a full development environment (via their choice of IDE, such as VSCode). P2 this found helpful:  \qt{it was also super helpful that no setup is required, [...] auto refreshes on saves and stuff: The ideal development situation that's already there}. 
A common slow loop in other systems is to have to walk through the study in its entirety to debug parts of the sequence, whereas our study browser supports quick sequence navigation and therein rapid stimuli iteration. 

ReVISitPy has its own unique feedback loops. When used to generate configurations, there may be a wide feedback loop between generating a configuration in a notebook, copying the generated configuration into a project, and running the project to view changes. To tighten this feedback loop, reVISitPy can preview the study within a notebook, ensuring the configuration is appropriate before leaving the notebook. However, additional exploration of desired affordances during the prototyping phase is useful future work.

\revised{An often overlooked interaction with study frameworks is with their documentation, examples, and tutorials. As part of our focus on community, we strive to have accessible documentation, along with examples covering a range of functionality (inspired by the sprawling D3 example gallery ecosystem~\cite{yang_considering_2024}), and descriptive tutorials for more complex concepts. 
For instance, P2 found our data storage setup tutorial helpful, saying that they 
\qt{100\% had to use the guide to be able to remember what to do}; however, \qt{I thought the guide was really good}. 
Similarly, P3 recalled that
\qt{I mainly learned by using the documents on the website and API documents.} 
No documentation is perfect, but we continue to adapt it to the needs of our user community.}

A key design decision was to task study designers with using a DSL to specify experiments, in contrast to a GUI (as Qualtrics does), which likely would have radically tightened the design loop. 
Although there is substantially higher technical complexity in using a DSL, it makes many more designs possible than what we might have designed for in building a GUI. Echoing Alan Kay: ``\emph{Simple things should be simple, complex things should be possible}''~\cite{alankay_alan}.
While future work might explore a GUI,
we suggest that centering the possible (and valuing the technical proficiencies of our design-for audience) is essential. 

\parahead{Notation} \textit{What notations are present and how do they interrelate?} ReVISit's primary notation is the JSON DSL (which every study interacts with), complemented by several optional ones. 
Our DSL is designed to be simple and highly composable, with the only syntactic abstraction being \texttt{baseComponents}, as indicated in \autoref{fig:grammar-fig}. 

Such simplicity can be misaligned with the expected simplicity of some experiment designs. As discussed in \autoref{section-dsl},
config files can be very large due to repetition of structures and components for a variety of factors.
P2, who needed to make a 12k LOC file, questioned \qt{I don't know if what I'm doing is just a really rare experimental design. It doesn't sound it from saying it in English, but then I think it kind of ends up being fairly orthogonal to the way the sequence is set up}.
ReVISitPy attempts to address this problem by offering a greater degree of abstraction than is present in the DSL.
The choice of transferring complexity away from the primary notation (the DSL) and into a secondary notation (reVISitPy/Python) was intended to keep the DSL simple, as most studies would not need the complex features that reVISitPy enables, while also empowering users who do need such functionalities. 
We forwent implementation of common abstractions (such as loops and variables) to avoid maintaining a full programming language. Instead, we ensure that our DSL is highly composable, allowing for complex designs via reVISitPy or other custom solutions, with the tradeoff of increased difficulty in generating the configuration.


Apart from the DSL, studies may utilize secondary notations for construction of stimuli---such as markdown or React, as well as any iframe embeddable content. 
Many stimuli interact with the DSL to access parameters, answers, or provenance data. 
To create custom stimuli, P1 used multiple notations, some of which they had no prior experience with. They pointed out that this increased the learning curve: \qt{You're also learning React and D3 and JavaScript and a little bit of Firebase}. The many possible notations that are usable in conjunction with reVISit widen the range of possible stimuli, but also result in a steeper learning curve and inherited complexity of such notations.

\parahead{Errors} \textit{What are they and how are they handled?}
Errors within reVISit are handled differently depending on the source of the error. Errors within our primary notation, the DSL, make rendering and compiling the study impossible, increasing the importance that study designers can quickly identify and fix these errors. Warnings about DSL errors can originate from two sources. The first is from automated validation of our JSON schema. The second is application-specific errors which the system throws, such as trying to use a \textit{baseComponent} which does not exist. In both cases, errors are immediately surfaced and shown as errors within the system with an error message. 
However, some errors may be difficult to identify via this message, especially errors that originated from schema validation. 
P3 noted, \qt{it would be better if that was a little bit more straightforward about exactly what part is broken}. 
However, P3 also pointed out the increased support that code editors can give to identify such errors, saying, \qt{I use VS Code for opening the JSON. And if there are grammar mistakes, then VS Code itself reveals which part is wrong. That was also helpful when the file was broken.} 
ReVISit surfaces lint-like warnings for stylistic usage problems---for example, components that are defined but not used yield a warning.
However, our linting is limited to programmatic problems in the DSL, and does not consider the semantics of an experimental design (\eg{} confounders). 
Although reVISit puts a number of experimental designs within reach, it does little to help ensure that the experimental designs themselves are good outside of these practices. For instance, an experiment may unintentionally introduce learning effects by always presenting a fixed order of chart stimuli that escalate in complexity. In future work, it would be useful to explore automated experiment evaluation and the effects it would have on experimental design. 

Errors that occur outside of the DSL can be more difficult to identify. ReVISit surfaces errors thrown by React components to avoid complete application crashes, but does not have such functionality for iframes. 
Errors can be surfaced in the development environment itself, as with an error P1 ran into: 
\qt{when I tried to import this [math library] into reVISit, it was telling me about some random number generating library that is outdated that revisit does not accept.} 
Errors such as this one, which often stem from library version conflicts, can be challenging to fix even for users with a background in web development. 
We could avoid such errors by dictating the development environment (\eg{} by allowing only some packages or using a GUI), however this is in tension with our experimental sovereignty goal: some things may be harder, but more things will be possible. 

\parahead{Customizability} \textit{How can programs be modified?}
%
Easy replication and modification of studies is central to reVISit's focus on improving reproducibility in user studies.
Our core GitHub repository comes with a collection of example studies, such as a replication of Cleveland-McGill~\cite{cleveland_graphical_1984}, in a manner that supports opportunistic programming-style reuse~\cite{brandt_two_2009}. 
Both the experiment design (via the DSL) and the stimuli are available and modifiable.
The same is true for public reVISit studies, such as those run by Lisnic et al~\cite{lisnic_visualization_2025}, which can be reproduced or modified easily. 
Studies maintain the version of reVISit when forked (unless manually merged), ensuring that the study continues to operate properly in spite of continued development on reVISit. 

Additionally, as study designers fork the entire codebase, core functionality of the tool is visible and changeable. 
P1 used the open nature of the core code as a learning tool, observing \qt{that's the benefit of having all the code base up there. If there's something that I don't understand, I can always just try to find its source}. 
P2 and P3 both took this a step further, editing the core code to match certain desired behavior for their studies, with P2 saying \qt{if I need to change something in its guts, I can do that, which is fun, as opposed to something more closed}. 
Lisnic \etal{}~\cite{lisnic_visualization_2025}, for example, modified the core code to host a ``\ourhref{https://vdl.sci.utah.edu/viz-guardrails-study/}{demonstration page}'' for their project, including the study, its results, and a sandboxed version of their stimulus.
Viewing or modifying core code was not originally intended behavior, and cannot be expected from study designers without a strong technical background. However, for those with the skills required, the core code can prove a valuable tool to accompany documentation and allow for a high level of customization. 




\parahead{Adaptability} \textit{What socio-technical (e.g., learnability) dimensions are considered?}
ReVISit is primarily directed at the visualization community, but has been designed with the intention of being usable for a wide range of studies, including from fields with less uniform technical backgrounds (\eg{} psychology). 
However, from our interviews (and experiences), reVISit is currently most useful for those with a programming background (inheriting learning curves from React or TypeScript).
Taking advantage of many of the unique and valuable features within reVISit, such as application provenance tracking or reVISitPy, requires programming knowledge.
Our choice to avoid GUI approaches (for the time being) is partially to blame for this learning curve. P3 pointed out that \qt{there are some people that do not have the computational knowledge of web designing. It's really difficult for me to suggest [to them that] they use revisit, because [..] you need to tune actual code, compared to Qualtrics or those web based survey systems}. 

One approach we have taken to mitigate the learning curve is via community outreach, which has been a focus since the first stable release. We have held tutorials on reVISit at VIS'24, \revised{CHI'25, EuroVis'25}, Georgia Tech, University of Utah, UNC Chapel Hill, and are \revised{scheduled to hold a tutorial at VIS'25}. We also have an active Slack team with public channels for users to get help and make suggestions. 

\section{Conclusion}


ReVISit 2 seeks to simplify the process of designing, debugging, deploying, and disseminating experiments in visualization research and related fields. 
Through a four year cycle of community engaged work we have developed a platform that we believe drastically simplifies the experimental process, especially for the kind of sophisticated studies that are increasingly commonplace in visualization and HCI.
As part of this work, we developed a range of novel contributions to experiment infrastructure, including a sophisticated DSL and a Python-based extension that can accomplish more diverse study designs than comparable frameworks, notebook-based prototypes,  provenance tracking and associated participant replay, and advanced stimuli, such as Vega and React programs.
To demonstrate these contributions (and the robustness of reVISit), we conducted a trio of replication studies, which we used (in tandem with a small interview study) to reflect on our design choices.

This work is not the end of reVISit's story. There are various additional improvements to the experiment life cycle to explore, new communities to support, and limitations to address. 
For instance, ReVISit is primarily designed for visualization research, and is focused by the particular research goals of the research team---although we made efforts to be expansive in our designs and be informed by a variety of stakeholders from around the visualization community. 
However, we do not address concerns held in related domains, such as games~\cite{partlan_exploratory_2018} or social robots~\cite{tsoi_approach_2021}. 
\revised{Similarly, reVISit is currently limited to desktop-based web browsers, precluding studies in mediums like mobile, watchfaces, or AR/VR. While many such studies can be replicated in web browsers, native support would be preferable.}

Through this work, and our continued work on reVISit, we hope to raise study quality and reproducibility for the visualization community and make it easier to investigate rich empirical phenomena.

\acknowledgments{
We would like to extend our thanks to the National Science Foundation (2213756, 2213757, 2402719),
to Carolina Nobre, to our users for their insightful questions and contributions, to our numerous study/interview participants, and to our community advisory board, which consists of Danielle Albers Szafir, Cindy Xiong Bearfield, Ana Crisan, Alex Endert, Jean-Daniel Fekete, Petra Isenberg, Lace Padilla, John Stasko, and Manuela Waldner.
}

\section*{Supplemental Materials}

The code for reVISit can be found at \ourhref{https://github.com/revisit-studies/study}{github.com/revisit-studies/study}. All documentation, examples, and tutorials can be found at \ourhref{https://revisit.dev/}{revisit.dev}. A demo with example studies is at {\ourhref{https://revisit.dev/study/}{revisit.dev/study}}. The study replication code is at \ourhref{https://github.com/revisit-studies/replication-studies}{github.com/revisit-studies/replication-studies}, and the full replication studies and their data are available at \ourhref{https://revisit.dev/replication-studies/}{revisit.dev/replication-studies}. We also provide an appendix with additional analysis for all replications, and a separate repository containing the data and the analysis code for the replications at \ourhref{https://github.com/revisit-studies/replication-studies-analysis}{github.com/revisit-studies/replication-studies-analysis}. Pre registrations for each study can be found at \ourhref{https://osf.io/e8anx/}{https://osf.io/e8anx/}.

\bibliographystyle{abbrv-doi-hyperref}
\bibliography{2025-revisit-2.bib}

\begin{thebibliography}{10}

\bibitem{abdalkareem_impact_2020}
R.~Abdalkareem, V.~Oda, S.~Mujahid, and E.~Shihab.
\newblock On the impact of using trivial packages: An empirical case study on npm and {{PyPI}}.
\newblock {\em Empirical Software Engineering}, 25(2):1168--1204, Mar. 2020. \href{https://doi.org/10.1007/s10664-019-09792-9}
{doi: {{%
10\hspace{.1pt}\discretionary{.}{%
}{.}\hspace{.4pt}1007\discretionary{/}{%
}{/}s10664\discretionary{%
}{-}{-}019\discretionary{%
}{-}{-}09792\discretionary{%
}{-}{-}9}}}


\bibitem{aigner_evalbench_2013}
W.~Aigner, S.~Hoffmann, and A.~Rind.
\newblock {{EvalBench}}: {{A Software Library}} for {{Visualization Evaluation}}.
\newblock {\em Computer Graphics Forum}, 32(3pt1):41--50, 2013. \href{https://doi.org/10.1111/cgf.12091}
{doi: {{%
10\hspace{.1pt}\discretionary{.}{%
}{.}\hspace{.4pt}1111\discretionary{/}{%
}{/}cgf\hspace{.1pt}\discretionary{.}{%
}{.}\hspace{.4pt}12091}}}


\bibitem{akbaba_troubling_2023}
D.~Akbaba, D.~Lange, M.~Correll, A.~Lex, and M.~Meyer.
\newblock Troubling {{Collaboration}}: {{Matters}} of {{Care}} for {{Visualization Design}}.
\newblock In {\em {{ACM CHI Conference}} on {{Human Factors}} in {{Computing Systems}}}, 2023. \href{https://doi.org/10.1145/3544548.3581168}
{doi: {{%
10\hspace{.1pt}\discretionary{.}{%
}{.}\hspace{.4pt}1145\discretionary{/}{%
}{/}3544548\hspace{.1pt}\discretionary{.}{%
}{.}\hspace{.4pt}3581168}}}


\bibitem{alankay_alan}
{Alan Kay}.
\newblock Alan {{Kay}}'s adage ``{{Simple}} things should be simple, complex things should be possible''.

\bibitem{boy_suggested_2016}
J.~Boy, L.~Eveillard, F.~Detienne, and J.-D. Fekete.
\newblock Suggested {{Interactivity}}: {{Seeking Perceived Affordances}} for {{Information Visualization}}.
\newblock {\em IEEE Transactions on Visualization and Computer Graphics}, 22(1):639--648, Jan. 2016. \href{https://doi.org/10.1109/TVCG.2015.2467201}
{doi: {{%
10\hspace{.1pt}\discretionary{.}{%
}{.}\hspace{.4pt}1109\discretionary{/}{%
}{/}TVCG\hspace{.1pt}\discretionary{.}{%
}{.}\hspace{.4pt}2015\hspace{.1pt}\discretionary{.}{%
}{.}\hspace{.4pt}2467201}}}


\bibitem{brandt_two_2009}
J.~Brandt, P.~J. Guo, J.~Lewenstein, M.~Dontcheva, and S.~R. Klemmer.
\newblock Two studies of opportunistic programming: Interleaving web foraging, learning, and writing code.
\newblock In {\em Proceedings of the {{SIGCHI Conference}} on {{Human Factors}} in {{Computing Systems}}}, pp. 1589--1598. ACM, Boston MA USA, Apr. 2009. \href{https://doi.org/10.1145/1518701.1518944}
{doi: {{%
10\hspace{.1pt}\discretionary{.}{%
}{.}\hspace{.4pt}1145\discretionary{/}{%
}{/}1518701\hspace{.1pt}\discretionary{.}{%
}{.}\hspace{.4pt}1518944}}}


\bibitem{chen_gazealytics_2023}
K.-T. Chen, A.~Prouzeau, J.~Langmead, R.~T. {Whitelock-Jones}, L.~Lawrence, T.~Dwyer, C.~Hurter, D.~Weiskopf, and S.~Goodwin.
\newblock Gazealytics: {{A Unified}} and {{Flexible Visual Toolkit}} for {{Exploratory}} and {{Comparative Gaze Analysis}}.
\newblock In {\em 2023 {{Symposium}} on {{Eye Tracking Research}} and {{Applications}}}, pp. 1--7, May 2023. \href{https://doi.org/10.1145/3588015.3589844}
{doi: {{%
10\hspace{.1pt}\discretionary{.}{%
}{.}\hspace{.4pt}1145\discretionary{/}{%
}{/}3588015\hspace{.1pt}\discretionary{.}{%
}{.}\hspace{.4pt}3589844}}}


\bibitem{cleveland_graphical_1984}
W.~S. Cleveland and R.~McGill.
\newblock Graphical {{Perception}}: {{Theory}}, {{Experimentation}}, and {{Application}} to the {{Development}} of {{Graphical Methods}}.
\newblock {\em Journal of the American Statistical Association}, 79(387):531--554, 1984. \href{https://doi.org/10.1080/01621459.1984.10478080}
{doi: {{%
10\hspace{.1pt}\discretionary{.}{%
}{.}\hspace{.4pt}1080\discretionary{/}{%
}{/}01621459\hspace{.1pt}\discretionary{.}{%
}{.}\hspace{.4pt}1984\hspace{.1pt}\discretionary{.}{%
}{.}\hspace{.4pt}10478080}}}


\bibitem{cohen_statistical_2013}
J.~Cohen.
\newblock {\em Statistical {{Power Analysis}} for the {{Behavioral Sciences}}}.
\newblock Routledge, May 2013.

\bibitem{cui_promises_2025}
Y.~Cui, L.~W. Ge, Y.~Ding, L.~Harrison, F.~Yang, and M.~Kay.
\newblock Promises and {{Pitfalls}}: {{Using Large Language Models}} to {{Generate Visualization Items}}.
\newblock {\em IEEE Transactions on Visualization and Computer Graphics}, 31(1):1094--1104, Jan. 2025. \href{https://doi.org/10.1109/TVCG.2024.3456309}
{doi: {{%
10\hspace{.1pt}\discretionary{.}{%
}{.}\hspace{.4pt}1109\discretionary{/}{%
}{/}TVCG\hspace{.1pt}\discretionary{.}{%
}{.}\hspace{.4pt}2024\hspace{.1pt}\discretionary{.}{%
}{.}\hspace{.4pt}3456309}}}


\bibitem{cutler_crowdsourced_2025}
Z.~Cutler, L.~Harrison, C.~Nobre, and A.~Lex.
\newblock Crowdsourced {{Think-Aloud Studies}}.
\newblock In {\em {{ACM Conference}} on {{Human Factors}} in {{Computing Systems}} ({{CHI}})}, 2025. \href{https://doi.org/10.1145/3706598.3714305}
{doi: {{%
10\hspace{.1pt}\discretionary{.}{%
}{.}\hspace{.4pt}1145\discretionary{/}{%
}{/}3706598\hspace{.1pt}\discretionary{.}{%
}{.}\hspace{.4pt}3714305}}}


\bibitem{cutler_trrack_2020}
Z.~T. Cutler, K.~Gadhave, and A.~Lex.
\newblock Trrack: {{A Library}} for {{Provenance Tracking}} in {{Web-Based Visualizations}}.
\newblock In {\em {{IEEE Visualization Conference}} ({{VIS}})}, pp. 116--120, 2020. \href{https://doi.org/10.1109/VIS47514.2020.00030}
{doi: {{%
10\hspace{.1pt}\discretionary{.}{%
}{.}\hspace{.4pt}1109\discretionary{/}{%
}{/}VIS47514\hspace{.1pt}\discretionary{.}{%
}{.}\hspace{.4pt}2020\hspace{.1pt}\discretionary{.}{%
}{.}\hspace{.4pt}00030}}}


\bibitem{leeuw_jspsych_2023}
J.~R. de~Leeuw, R.~A. Gilbert, and B.~Luchterhandt.
\newblock {{jsPsych}}: {{Enabling}} an {{Open-Source Collaborative Ecosystem}} of {{Behavioral Experiments}}.
\newblock {\em Journal of Open Source Software}, 8(85):5351, May 2023. \href{https://doi.org/10.21105/joss.05351}
{doi: {{%
10\hspace{.1pt}\discretionary{.}{%
}{.}\hspace{.4pt}21105\discretionary{/}{%
}{/}joss\hspace{.1pt}\discretionary{.}{%
}{.}\hspace{.4pt}05351}}}


\bibitem{ding_revisit_2023}
Y.~Ding, J.~Wilburn, H.~Shrestha, A.~Ndlovu, K.~Gadhave, C.~Nobre, A.~Lex, and L.~Harrison.
\newblock {{reVISit}}: {{Supporting Scalable Evaluation}} of {{Interactive Visualizations}}.
\newblock In {\em {{IEEE Visualization}} and {{Visual Analytics}} ({{VIS}})}, pp. 31--35. IEEE, 2023. \href{https://doi.org/10.1109/VIS54172.2023.00015}
{doi: {{%
10\hspace{.1pt}\discretionary{.}{%
}{.}\hspace{.4pt}1109\discretionary{/}{%
}{/}VIS54172\hspace{.1pt}\discretionary{.}{%
}{.}\hspace{.4pt}2023\hspace{.1pt}\discretionary{.}{%
}{.}\hspace{.4pt}00015}}}


\bibitem{dou_comparing_2010}
W.~Dou, C.~Ziemkiewicz, L.~Harrison, D.~H. Jeong, R.~Ryan, W.~Ribarsky, X.~Wang, and R.~Chang.
\newblock Comparing different levels of interaction constraints for deriving visual problem isomorphs.
\newblock In {\em 2010 {{IEEE Symposium}} on {{Visual Analytics Science}} and {{Technology}}}, pp. 195--202, Oct. 2010. \href{https://doi.org/10.1109/VAST.2010.5653599}
{doi: {{%
10\hspace{.1pt}\discretionary{.}{%
}{.}\hspace{.4pt}1109\discretionary{/}{%
}{/}VAST\hspace{.1pt}\discretionary{.}{%
}{.}\hspace{.4pt}2010\hspace{.1pt}\discretionary{.}{%
}{.}\hspace{.4pt}5653599}}}


\bibitem{eiselmayer_touchstone2_2019}
A.~Eiselmayer, C.~Wacharamanotham, M.~{Beaudouin-Lafon}, and W.~E. Mackay.
\newblock Touchstone2: {{An Interactive Environment}} for {{Exploring Trade-offs}} in {{HCI Experiment Design}}.
\newblock In {\em Proceedings of the 2019 {{CHI Conference}} on {{Human Factors}} in {{Computing Systems}}}, pp. 1--11. ACM, Glasgow Scotland Uk, May 2019. \href{https://doi.org/10.1145/3290605.3300447}
{doi: {{%
10\hspace{.1pt}\discretionary{.}{%
}{.}\hspace{.4pt}1145\discretionary{/}{%
}{/}3290605\hspace{.1pt}\discretionary{.}{%
}{.}\hspace{.4pt}3300447}}}


\bibitem{fan_vista_2020}
M.~Fan, K.~Wu, J.~Zhao, Y.~Li, W.~Wei, and K.~N. Truong.
\newblock {{VisTA}}: {{Integrating Machine Intelligence}} with {{Visualization}} to {{Support}} the {{Investigation}} of {{Think-Aloud Sessions}}.
\newblock {\em IEEE Transactions on Visualization and Computer Graphics}, 26(1):343--352, 2020. \href{https://doi.org/10.1109/TVCG.2019.2934797}
{doi: {{%
10\hspace{.1pt}\discretionary{.}{%
}{.}\hspace{.4pt}1109\discretionary{/}{%
}{/}TVCG\hspace{.1pt}\discretionary{.}{%
}{.}\hspace{.4pt}2019\hspace{.1pt}\discretionary{.}{%
}{.}\hspace{.4pt}2934797}}}


\bibitem{feng_effects_2018}
M.~Feng, C.~Deng, E.~M. Peck, and L.~Harrison.
\newblock The {{Effects}} of {{Adding Search Functionality}} to {{Interactive Visualizations}} on the {{Web}}.
\newblock In {\em Proceedings of the {{SIGCHI Conference}} on {{Human Factors}} in {{Computing Systems}} ({{CHI}})}, pp. 1--13. ACM, 2018. \href{https://doi.org/10.1145/3173574.3173711}
{doi: {{%
10\hspace{.1pt}\discretionary{.}{%
}{.}\hspace{.4pt}1145\discretionary{/}{%
}{/}3173574\hspace{.1pt}\discretionary{.}{%
}{.}\hspace{.4pt}3173711}}}


\bibitem{gergle_experimental_2014}
D.~Gergle and D.~S. Tan.
\newblock Experimental {{Research}} in {{HCI}}.
\newblock In J.~S. Olson and W.~A. Kellogg, eds., {\em Ways of {{Knowing}} in {{HCI}}}, pp. 191--227. Springer, New York, NY, 2014. \href{https://doi.org/10.1007/978-1-4939-0378-8_9}
{doi: {{%
10\hspace{.1pt}\discretionary{.}{%
}{.}\hspace{.4pt}1007\discretionary{/}{%
}{/}978\discretionary{%
}{-}{-}1\discretionary{%
}{-}{-}4939\discretionary{%
}{-}{-}0378\discretionary{%
}{-}{-}8\_9}}}


\bibitem{ghoniem_readability_2005}
M.~Ghoniem, J.-D. Fekete, and P.~Castagliola.
\newblock On the {{Readability}} of {{Graphs Using Node-Link}} and {{Matrix-Based Representations}}: {{A Controlled Experiment}} and {{Statistical Analysis}}.
\newblock {\em Information Visualization}, 4(2):114--135, 2005. \href{https://doi.org/10.1057/palgrave.ivs.9500092}
{doi: {{%
10\hspace{.1pt}\discretionary{.}{%
}{.}\hspace{.4pt}1057\discretionary{/}{%
}{/}palgrave\hspace{.1pt}\discretionary{.}{%
}{.}\hspace{.4pt}ivs\hspace{.1pt}\discretionary{.}{%
}{.}\hspace{.4pt}9500092}}}


\bibitem{google_google_2025}
{Google}.
\newblock Google {{Forms}}, Mar. 2025.

\bibitem{green_cognitive_1989}
T.~R. Green.
\newblock Cognitive dimensions of notations.
\newblock {\em People and computers V}, pp. 443--460, 1989. \href{https://doi.org/10.1007/3-540-44617-6_31}
{doi: {{%
10\hspace{.1pt}\discretionary{.}{%
}{.}\hspace{.4pt}1007\discretionary{/}{%
}{/}3\discretionary{%
}{-}{-}540\discretionary{%
}{-}{-}44617\discretionary{%
}{-}{-}6\_31}}}


\bibitem{harrison_experimentr_2019}
L.~Harrison, C.~Gramazio, F.~Y{\'a}ng, K.~Aragam, E.~Peck, and D.~Schroeder.
\newblock Experimentr, Feb. 2019.

\bibitem{harrison_ranking_2014}
L.~Harrison, F.~Yang, S.~Franconeri, and R.~Chang.
\newblock Ranking {{Visualizations}} of {{Correlation Using Weber}}'s {{Law}}.
\newblock {\em IEEE Transactions on Visualization and Computer Graphics}, 20(12):1943--1952, Dec. 2014. \href{https://doi.org/10.1109/TVCG.2014.2346979}
{doi: {{%
10\hspace{.1pt}\discretionary{.}{%
}{.}\hspace{.4pt}1109\discretionary{/}{%
}{/}TVCG\hspace{.1pt}\discretionary{.}{%
}{.}\hspace{.4pt}2014\hspace{.1pt}\discretionary{.}{%
}{.}\hspace{.4pt}2346979}}}


\bibitem{he_beauvis_2023}
T.~He, P.~Isenberg, R.~Dachselt, and T.~Isenberg.
\newblock {{BeauVis}}: {{A Validated Scale}} for {{Measuring}} the {{Aesthetic Pleasure}} of {{Visual Representations}}.
\newblock {\em IEEE Transactions on Visualization and Computer Graphics}, 29(1):363--373, Jan. 2023. \href{https://doi.org/10.1109/TVCG.2022.3209390}
{doi: {{%
10\hspace{.1pt}\discretionary{.}{%
}{.}\hspace{.4pt}1109\discretionary{/}{%
}{/}TVCG\hspace{.1pt}\discretionary{.}{%
}{.}\hspace{.4pt}2022\hspace{.1pt}\discretionary{.}{%
}{.}\hspace{.4pt}3209390}}}


\bibitem{he_design_2024}
T.~He, Y.~Zhong, P.~Isenberg, and T.~Isenberg.
\newblock Design {{Characterization}} for {{Black-and-White Textures}} in {{Visualization}}.
\newblock {\em IEEE Transactions on Visualization and Computer Graphics}, 30(1):1019--1029, Jan. 2024. \href{https://doi.org/10.1109/TVCG.2023.3326941}
{doi: {{%
10\hspace{.1pt}\discretionary{.}{%
}{.}\hspace{.4pt}1109\discretionary{/}{%
}{/}TVCG\hspace{.1pt}\discretionary{.}{%
}{.}\hspace{.4pt}2023\hspace{.1pt}\discretionary{.}{%
}{.}\hspace{.4pt}3326941}}}


\bibitem{heer_crowdsourcing_2010}
J.~Heer and M.~Bostock.
\newblock Crowdsourcing graphical perception: Using mechanical turk to assess visualization design.
\newblock In {\em Proceedings of the {{SIGCHI Conference}} on {{Human Factors}} in {{Computing Systems}} ({{CHI}})}, pp. 203--212. ACM, 2010. \href{https://doi.org/10.1145/1753326.1753357}
{doi: {{%
10\hspace{.1pt}\discretionary{.}{%
}{.}\hspace{.4pt}1145\discretionary{/}{%
}{/}1753326\hspace{.1pt}\discretionary{.}{%
}{.}\hspace{.4pt}1753357}}}


\bibitem{amazoninc._amazon_2025}
A.~Inc.
\newblock Amazon {{Mechanical Turk}}, 2025.

\bibitem{qualtricsinternationalinc._qualtrics_2025}
Q.~I. Inc.
\newblock Qualtrics, 2025.

\bibitem{jakubovic_technical_2023}
J.~Jakubovic, J.~Edwards, and T.~Petricek.
\newblock Technical {{Dimensions}} of {{Programming Systems}}.
\newblock {\em The Art, Science, and Engineering of Programming}, 7(3):13, 2023. \href{https://doi.org/10.22152/programming-journal.org/2023/7/13}
{doi: {{%
10\hspace{.1pt}\discretionary{.}{%
}{.}\hspace{.4pt}22152\discretionary{/}{%
}{/}programming\discretionary{%
}{-}{-}journal\hspace{.1pt}\discretionary{.}{%
}{.}\hspace{.4pt}org\discretionary{/}{%
}{/}2023\discretionary{/}{%
}{/}7\discretionary{/}{%
}{/}13}}}


\bibitem{jansen_froe_2025}
Y.~Jansen.
\newblock {{FROE}}: {{Framework}} for {{Running Online Experiments}}, Mar. 2025.

\bibitem{jeremyashkenas_how_2014}
{Jeremy Ashkenas} and {Alicia Parlapiano}.
\newblock How the {{Recession Reshaped}} the {{Economy}}, in 255 {{Charts}} - {{The New York Times}}, June 2014.

\bibitem{jianu_visunit_2025}
R.~Jianu, D.~Laksono, A.~Slingsby, and M.~Okoe.
\newblock {{VisUnit}}: {{Literate Visualisation Studies Assembled}} from {{Reusable Test-Suites}}.
\newblock In {\em {{ACM CHI Conference}} on {{Human Factors}} in {{Computing Systems}}}, 2025. \href{https://doi.org/10.1145/3706598.3713104}
{doi: {{%
10\hspace{.1pt}\discretionary{.}{%
}{.}\hspace{.4pt}1145\discretionary{/}{%
}{/}3706598\hspace{.1pt}\discretionary{.}{%
}{.}\hspace{.4pt}3713104}}}


\bibitem{joshi_likert_2015}
A.~Joshi, S.~Kale, S.~Chandel, and D.~Pal.
\newblock Likert {{Scale}}: {{Explored}} and {{Explained}}.
\newblock {\em British Journal of Applied Science \& Technology}, 7(4):396--403, Jan. 2015. \href{https://doi.org/10.9734/BJAST/2015/14975}
{doi: {{%
10\hspace{.1pt}\discretionary{.}{%
}{.}\hspace{.4pt}9734\discretionary{/}{%
}{/}BJAST\discretionary{/}{%
}{/}2015\discretionary{/}{%
}{/}14975}}}


\bibitem{kale_hypothetical_2018}
A.~Kale, F.~Nguyen, M.~Kay, and J.~Hullman.
\newblock Hypothetical outcome plots help untrained observers judge trends in ambiguous data.
\newblock {\em IEEE transactions on visualization and computer graphics}, 25(1):892--902, 2018. \href{https://doi.org/10.1109/TVCG.2018.2864909}
{doi: {{%
10\hspace{.1pt}\discretionary{.}{%
}{.}\hspace{.4pt}1109\discretionary{/}{%
}{/}TVCG\hspace{.1pt}\discretionary{.}{%
}{.}\hspace{.4pt}2018\hspace{.1pt}\discretionary{.}{%
}{.}\hspace{.4pt}2864909}}}


\bibitem{kay_webers_2016}
M.~Kay and J.~Heer.
\newblock Beyond {{Weber}}'s {{Law}}: {{A Second Look}} at {{Ranking Visualizations}} of {{Correlation}}.
\newblock {\em IEEE Transactions on Visualization and Computer Graphics}, 22(1):469--478, Jan. 2016. \href{https://doi.org/10.1109/TVCG.2015.2467671}
{doi: {{%
10\hspace{.1pt}\discretionary{.}{%
}{.}\hspace{.4pt}1109\discretionary{/}{%
}{/}TVCG\hspace{.1pt}\discretionary{.}{%
}{.}\hspace{.4pt}2015\hspace{.1pt}\discretionary{.}{%
}{.}\hspace{.4pt}2467671}}}


\bibitem{kieffer_hola_2016}
S.~Kieffer, T.~Dwyer, K.~Marriott, and M.~Wybrow.
\newblock {{HOLA}}: {{Human-like Orthogonal Network Layout}}.
\newblock {\em IEEE Transactions on Visualization and Computer Graphics}, 22(1):349--358, Jan. 2016. \href{https://doi.org/10.1109/TVCG.2015.2467451}
{doi: {{%
10\hspace{.1pt}\discretionary{.}{%
}{.}\hspace{.4pt}1109\discretionary{/}{%
}{/}TVCG\hspace{.1pt}\discretionary{.}{%
}{.}\hspace{.4pt}2015\hspace{.1pt}\discretionary{.}{%
}{.}\hspace{.4pt}2467451}}}


\bibitem{kim_designing_2019}
Y.~Kim, M.~Correll, and J.~Heer.
\newblock Designing {{Animated Transitions}} to {{Convey Aggregate Operations}}.
\newblock {\em Computer Graphics Forum}, 38(3):541--551, June 2019. \href{https://doi.org/10.1111/cgf.13709}
{doi: {{%
10\hspace{.1pt}\discretionary{.}{%
}{.}\hspace{.4pt}1111\discretionary{/}{%
}{/}cgf\hspace{.1pt}\discretionary{.}{%
}{.}\hspace{.4pt}13709}}}


\bibitem{lam_concept_2024}
M.~S. Lam, J.~Teoh, J.~A. Landay, J.~Heer, and M.~S. Bernstein.
\newblock Concept {{Induction}}: {{Analyzing Unstructured Text}} with {{High-Level Concepts Using LLooM}}.
\newblock In {\em Proceedings of the {{CHI Conference}} on {{Human Factors}} in {{Computing Systems}}}, {{CHI}} '24, pp. 1--28. Association for Computing Machinery, New York, NY, USA, May 2024. \href{https://doi.org/10.1145/3613904.3642830}
{doi: {{%
10\hspace{.1pt}\discretionary{.}{%
}{.}\hspace{.4pt}1145\discretionary{/}{%
}{/}3613904\hspace{.1pt}\discretionary{.}{%
}{.}\hspace{.4pt}3642830}}}


\bibitem{lee_how_2016}
S.~Lee, S.-H. Kim, Y.-H. Hung, H.~Lam, Y.-A. Kang, and J.~S. Yi.
\newblock How do {{People Make Sense}} of {{Unfamiliar Visualizations}}?: {{A Grounded Model}} of {{Novice}}'s {{Information Visualization Sensemaking}}.
\newblock {\em IEEE Transactions on Visualization and Computer Graphics}, 22(1):499--508, Jan. 2016. \href{https://doi.org/10.1109/TVCG.2015.2467195}
{doi: {{%
10\hspace{.1pt}\discretionary{.}{%
}{.}\hspace{.4pt}1109\discretionary{/}{%
}{/}TVCG\hspace{.1pt}\discretionary{.}{%
}{.}\hspace{.4pt}2015\hspace{.1pt}\discretionary{.}{%
}{.}\hspace{.4pt}2467195}}}


\bibitem{lee_vlat_2017}
S.~Lee, S.-H. Kim, and B.~C. Kwon.
\newblock {{VLAT}}: {{Development}} of a {{Visualization Literacy Assessment Test}}.
\newblock {\em IEEE Transactions on Visualization and Computer Graphics}, 23(1):551--560, 2017. \href{https://doi.org/10.1109/TVCG.2016.2598920}
{doi: {{%
10\hspace{.1pt}\discretionary{.}{%
}{.}\hspace{.4pt}1109\discretionary{/}{%
}{/}TVCG\hspace{.1pt}\discretionary{.}{%
}{.}\hspace{.4pt}2016\hspace{.1pt}\discretionary{.}{%
}{.}\hspace{.4pt}2598920}}}


\bibitem{lisnic_visualization_2025}
M.~Lisnic, Z.~Cutler, M.~Kogan, and A.~Lex.
\newblock Visualization guardrails: {{Designing}} interventions against cherry-picking in interactive data explorers.
\newblock In {\em {{SIGCHI}} Conference on Human Factors in Computing Systems ({{CHI}})}, 2025. \href{https://doi.org/10.1145/3706598.3713385}
{doi: {{%
10\hspace{.1pt}\discretionary{.}{%
}{.}\hspace{.4pt}1145\discretionary{/}{%
}{/}3706598\hspace{.1pt}\discretionary{.}{%
}{.}\hspace{.4pt}3713385}}}


\bibitem{lobo_flex-er_2020}
M.~J. Lobo, C.~Hurter, and P.~Irani.
\newblock Flex-{{ER}}: {{A Platform}} to {{Evaluate Interaction Techniques}} for {{Immersive Visualizations}}.
\newblock {\em Proc. ACM Hum.-Comput. Interact.}, 4(ISS):195:1--195:20, Nov. 2020. \href{https://doi.org/10.1145/3427323}
{doi: {{%
10\hspace{.1pt}\discretionary{.}{%
}{.}\hspace{.4pt}1145\discretionary{/}{%
}{/}3427323}}}


\bibitem{lyi_learnable_2024}
S.~L'Yi, A.~{van den Brandt}, E.~Adams, H.~N. Nguyen, and N.~Gehlenborg.
\newblock Learnable and {{Expressive Visualization Authoring}} through {{Blended Interfaces}}.
\newblock {\em IEEE Transactions on Visualization and Computer Graphics}, 2024. \href{https://doi.org/10.1109/TVCG.2024.3456598}
{doi: {{%
10\hspace{.1pt}\discretionary{.}{%
}{.}\hspace{.4pt}1109\discretionary{/}{%
}{/}TVCG\hspace{.1pt}\discretionary{.}{%
}{.}\hspace{.4pt}2024\hspace{.1pt}\discretionary{.}{%
}{.}\hspace{.4pt}3456598}}}


\bibitem{mackay_touchstone_2007}
W.~E. Mackay, C.~Appert, M.~{Beaudouin-Lafon}, O.~Chapuis, Y.~Du, J.-D. Fekete, and Y.~Guiard.
\newblock Touchstone: Exploratory design of experiments.
\newblock In {\em Proceedings of the {{SIGCHI Conference}} on {{Human Factors}} in {{Computing Systems}}}, {{CHI}} '07, pp. 1425--1434. Association for Computing Machinery, New York, NY, USA, Apr. 2007. \href{https://doi.org/10.1145/1240624.1240840}
{doi: {{%
10\hspace{.1pt}\discretionary{.}{%
}{.}\hspace{.4pt}1145\discretionary{/}{%
}{/}1240624\hspace{.1pt}\discretionary{.}{%
}{.}\hspace{.4pt}1240840}}}


\bibitem{mcnutt_accessible_2025}
A.~McNutt, M.~K. McCracken, I.~J. Eliza, D.~Hajas, J.~Wagoner, N.~Lanza, J.~Wilburn, S.~{Creem-Regehr}, and A.~Lex.
\newblock Accessible {{Text Descriptions}} for {{UpSet Plots}}.
\newblock {\em Computer Graphics Forum}, 44(3):e70102, 2025. \href{https://doi.org/10.1111/cgf.70102}
{doi: {{%
10\hspace{.1pt}\discretionary{.}{%
}{.}\hspace{.4pt}1111\discretionary{/}{%
}{/}cgf\hspace{.1pt}\discretionary{.}{%
}{.}\hspace{.4pt}70102}}}


\bibitem{mcnutt_mixing_2025}
A.~McNutt, M.~C. Stone, and J.~Heer.
\newblock Mixing {{Linters}} with {{GUIs}}: {{A Color Palette Design Probe}}.
\newblock {\em IEEE Transactions on Visualization and Computer Graphics}, 31(1):327--337, 2025. \href{https://doi.org/10.1109/TVCG.2024.3456317}
{doi: {{%
10\hspace{.1pt}\discretionary{.}{%
}{.}\hspace{.4pt}1109\discretionary{/}{%
}{/}TVCG\hspace{.1pt}\discretionary{.}{%
}{.}\hspace{.4pt}2024\hspace{.1pt}\discretionary{.}{%
}{.}\hspace{.4pt}3456317}}}


\bibitem{mcnutt_no_2023}
A.~M. McNutt.
\newblock No {{Grammar}} to {{Rule Them All}}: {{A Survey}} of {{JSON-style DSLs}} for {{Visualization}}.
\newblock {\em IEEE Transactions on Visualization and Computer Graphics}, 29(1):160--170, 2023. \href{https://doi.org/10.1109/TVCG.2022.3209460}
{doi: {{%
10\hspace{.1pt}\discretionary{.}{%
}{.}\hspace{.4pt}1109\discretionary{/}{%
}{/}TVCG\hspace{.1pt}\discretionary{.}{%
}{.}\hspace{.4pt}2022\hspace{.1pt}\discretionary{.}{%
}{.}\hspace{.4pt}3209460}}}


\bibitem{meuschke_evalviz_2019}
M.~Meuschke, N.~N. Smit, N.~Lichtenberg, B.~Preim, and K.~Lawonn.
\newblock {{EvalViz}} -- {{Surface}} visualization evaluation wizard for depth and shape perception tasks.
\newblock {\em Computers \& Graphics}, 82:250--263, Aug. 2019. \href{https://doi.org/10.1016/j.cag.2019.05.022}
{doi: {{%
10\hspace{.1pt}\discretionary{.}{%
}{.}\hspace{.4pt}1016\discretionary{/}{%
}{/}j\hspace{.1pt}\discretionary{.}{%
}{.}\hspace{.4pt}cag\hspace{.1pt}\discretionary{.}{%
}{.}\hspace{.4pt}2019\hspace{.1pt}\discretionary{.}{%
}{.}\hspace{.4pt}05\hspace{.1pt}\discretionary{.}{%
}{.}\hspace{.4pt}022}}}


\bibitem{musslick_sweetpea_2022}
S.~Musslick, A.~Cherkaev, B.~Draut, A.~S. Butt, P.~Darragh, V.~Srikumar, M.~Flatt, and J.~D. Cohen.
\newblock {{SweetPea}}: {{A}} standard language for factorial experimental design.
\newblock {\em Behavior Research Methods}, 54(2):805--829, Apr. 2022. \href{https://doi.org/10.3758/s13428-021-01598-2}
{doi: {{%
10\hspace{.1pt}\discretionary{.}{%
}{.}\hspace{.4pt}3758\discretionary{/}{%
}{/}s13428\discretionary{%
}{-}{-}021\discretionary{%
}{-}{-}01598\discretionary{%
}{-}{-}2}}}


\bibitem{nobre_revisit_2021}
C.~Nobre, D.~Wootton, Z.~Cutler, L.~Harrison, H.~Pfister, and A.~Lex.
\newblock {{reVISit}}: {{Looking Under}} the {{Hood}} of {{Interactive Visualization Studies}}.
\newblock In {\em {{SIGCHI Conference}} on {{Human Factors}} in {{Computing Systems}}}, pp. 1--13. ACM, Yokohama Japan, May 2021. \href{https://doi.org/10.1145/3411764.3445382}
{doi: {{%
10\hspace{.1pt}\discretionary{.}{%
}{.}\hspace{.4pt}1145\discretionary{/}{%
}{/}3411764\hspace{.1pt}\discretionary{.}{%
}{.}\hspace{.4pt}3445382}}}


\bibitem{nobre_evaluating_2020}
C.~Nobre, D.~Wootton, L.~Harrison, and A.~Lex.
\newblock Evaluating {{Multivariate Network Visualization Techniques Using}} a {{Validated Design}} and {{Crowdsourcing Approach}}.
\newblock In {\em {{SIGCHI Conference}} on {{Human Factors}} in {{Computing Systems}}}, pp. 1--12. ACM, 2020. \href{https://doi.org/10.1145/3313831.3376381}
{doi: {{%
10\hspace{.1pt}\discretionary{.}{%
}{.}\hspace{.4pt}1145\discretionary{/}{%
}{/}3313831\hspace{.1pt}\discretionary{.}{%
}{.}\hspace{.4pt}3376381}}}


\bibitem{nowak_designing_2024}
S.~Nowak and L.~Bartram.
\newblock Designing for {{Ambiguity}} in {{Visual Analytics}}: {{Lessons}} from {{Risk Assessment}} and {{Prediction}}.
\newblock {\em IEEE Transactions on Visualization and Computer Graphics}, 30(1):924--933, Jan. 2024. \href{https://doi.org/10.1109/TVCG.2023.3326571}
{doi: {{%
10\hspace{.1pt}\discretionary{.}{%
}{.}\hspace{.4pt}1109\discretionary{/}{%
}{/}TVCG\hspace{.1pt}\discretionary{.}{%
}{.}\hspace{.4pt}2023\hspace{.1pt}\discretionary{.}{%
}{.}\hspace{.4pt}3326571}}}


\bibitem{okoe_graphunit_2015}
M.~Okoe and R.~Jianu.
\newblock {{GraphUnit}}: {{Evaluating Interactive Graph Visualizations Using Crowdsourcing}}.
\newblock {\em Computer Graphics Forum}, 34(3):451--460, 2015. \href{https://doi.org/10.1111/cgf.12657}
{doi: {{%
10\hspace{.1pt}\discretionary{.}{%
}{.}\hspace{.4pt}1111\discretionary{/}{%
}{/}cgf\hspace{.1pt}\discretionary{.}{%
}{.}\hspace{.4pt}12657}}}


\bibitem{palan_prolific.ac_2018}
S.~Palan and C.~Schitter.
\newblock Prolific.ac {{A}} subject pool for online experiments.
\newblock {\em Journal of Behavioral and Experimental Finance}, 17:22--27, Mar. 2018. \href{https://doi.org/10.1016/j.jbef.2017.12.004}
{doi: {{%
10\hspace{.1pt}\discretionary{.}{%
}{.}\hspace{.4pt}1016\discretionary{/}{%
}{/}j\hspace{.1pt}\discretionary{.}{%
}{.}\hspace{.4pt}jbef\hspace{.1pt}\discretionary{.}{%
}{.}\hspace{.4pt}2017\hspace{.1pt}\discretionary{.}{%
}{.}\hspace{.4pt}12\hspace{.1pt}\discretionary{.}{%
}{.}\hspace{.4pt}004}}}


\bibitem{pandey_mini-vlat_2023}
S.~Pandey and A.~Ottley.
\newblock Mini-{{VLAT}}: {{A Short}} and {{Effective Measure}} of {{Visualization Literacy}}.
\newblock {\em Computer Graphics Forum}, 42(3):1--11, 2023. \href{https://doi.org/10.1111/cgf.14809}
{doi: {{%
10\hspace{.1pt}\discretionary{.}{%
}{.}\hspace{.4pt}1111\discretionary{/}{%
}{/}cgf\hspace{.1pt}\discretionary{.}{%
}{.}\hspace{.4pt}14809}}}


\bibitem{partlan_exploratory_2018}
N.~Partlan, E.~Carstensdottir, S.~Snodgrass, E.~Kleinman, G.~Smith, C.~Harteveld, and M.~S. {El-Nasr}.
\newblock Exploratory automated analysis of structural features of interactive narrative.
\newblock In {\em Proceedings of the {{AAAI Conference}} on {{Artificial Intelligence}} and {{Interactive Digital Entertainment}}}, vol.~14, pp. 88--94, 2018. \href{https://doi.org/10.1609/aiide.v14i1.13019}
{doi: {{%
10\hspace{.1pt}\discretionary{.}{%
}{.}\hspace{.4pt}1609\discretionary{/}{%
}{/}aiide\hspace{.1pt}\discretionary{.}{%
}{.}\hspace{.4pt}v14i1\hspace{.1pt}\discretionary{.}{%
}{.}\hspace{.4pt}13019}}}


\bibitem{peirce_psychopy2_2019}
J.~Peirce, J.~R. Gray, S.~Simpson, M.~MacAskill, R.~H{\"o}chenberger, H.~Sogo, E.~Kastman, and J.~K. Lindel{\o}v.
\newblock {{PsychoPy2}}: {{Experiments}} in behavior made easy.
\newblock {\em Behavior Research Methods}, 51(1):195--203, Feb. 2019. \href{https://doi.org/10.3758/s13428-018-01193-y}
{doi: {{%
10\hspace{.1pt}\discretionary{.}{%
}{.}\hspace{.4pt}3758\discretionary{/}{%
}{/}s13428\discretionary{%
}{-}{-}018\discretionary{%
}{-}{-}01193\discretionary{%
}{-}{-}y}}}


\bibitem{reinecke_labinthewild_2015}
K.~Reinecke and K.~Z. Gajos.
\newblock {{LabintheWild}}: {{Conducting Large-Scale Online Experiments With Uncompensated Samples}}.
\newblock In {\em Proceedings of the 18th {{ACM Conference}} on {{Computer Supported Cooperative Work}} \& {{Social Computing}}}, {{CSCW}}, pp. 1364--1378. Association for Computing Machinery, New York, 2015. \href{https://doi.org/10.1145/2675133.2675246}
{doi: {{%
10\hspace{.1pt}\discretionary{.}{%
}{.}\hspace{.4pt}1145\discretionary{/}{%
}{/}2675133\hspace{.1pt}\discretionary{.}{%
}{.}\hspace{.4pt}2675246}}}


\bibitem{satyanarayan_vega-lite_2017}
A.~Satyanarayan, D.~Moritz, K.~Wongsuphasawat, and J.~Heer.
\newblock Vega-{{Lite}}: {{A Grammar}} of {{Interactive Graphics}}.
\newblock {\em IEEE Transactions on Visualization and Computer Graphics}, 23(1):341--350, Jan. 2017. \href{https://doi.org/10.1109/TVCG.2016.2599030}
{doi: {{%
10\hspace{.1pt}\discretionary{.}{%
}{.}\hspace{.4pt}1109\discretionary{/}{%
}{/}TVCG\hspace{.1pt}\discretionary{.}{%
}{.}\hspace{.4pt}2016\hspace{.1pt}\discretionary{.}{%
}{.}\hspace{.4pt}2599030}}}


\bibitem{soure_coux_2022}
E.~J. Soure, E.~Kuang, M.~Fan, and J.~Zhao.
\newblock {{CoUX}}: {{Collaborative Visual Analysis}} of {{Think-Aloud Usability Test Videos}} for {{Digital Interfaces}}.
\newblock {\em IEEE Transactions on Visualization and Computer Graphics}, 28(1):643--653, Jan. 2022. \href{https://doi.org/10.1109/TVCG.2021.3114822}
{doi: {{%
10\hspace{.1pt}\discretionary{.}{%
}{.}\hspace{.4pt}1109\discretionary{/}{%
}{/}TVCG\hspace{.1pt}\discretionary{.}{%
}{.}\hspace{.4pt}2021\hspace{.1pt}\discretionary{.}{%
}{.}\hspace{.4pt}3114822}}}


\bibitem{stoet_psytoolkit_2017}
G.~Stoet.
\newblock {{PsyToolkit}}: {{A Novel Web-Based Method}} for {{Running Online Questionnaires}} and {{Reaction-Time Experiments}}.
\newblock {\em Teaching of Psychology}, 44(1):24--31, Jan. 2017. \href{https://doi.org/10.1177/0098628316677643}
{doi: {{%
10\hspace{.1pt}\discretionary{.}{%
}{.}\hspace{.4pt}1177\discretionary{/}{%
}{/}0098628316677643}}}


\bibitem{sukumar_designing_2018}
P.~T. Sukumar and R.~Metoyer.
\newblock Towards {{Designing Unbiased Replication Studies}} in {{Information Visualization}}.
\newblock In {\em 2018 {{IEEE Evaluation}} and {{Beyond}} - {{Methodological Approaches}} for {{Visualization}} ({{BELIV}})}, pp. 93--101, 2018. \href{https://doi.org/10.1109/BELIV.2018.8634261}
{doi: {{%
10\hspace{.1pt}\discretionary{.}{%
}{.}\hspace{.4pt}1109\discretionary{/}{%
}{/}BELIV\hspace{.1pt}\discretionary{.}{%
}{.}\hspace{.4pt}2018\hspace{.1pt}\discretionary{.}{%
}{.}\hspace{.4pt}8634261}}}


\bibitem{surveymonkey_surveymonkey_2025}
{SurveyMonkey}.
\newblock {{SurveyMonkey}}, Mar. 2025.

\bibitem{szafir_modeling_2018}
D.~A. Szafir.
\newblock Modeling {{Color Difference}} for {{Visualization Design}}.
\newblock {\em IEEE Transactions on Visualization and Computer Graphics}, 24(1):392--401, Jan. 2018. \href{https://doi.org/10.1109/TVCG.2017.2744359}
{doi: {{%
10\hspace{.1pt}\discretionary{.}{%
}{.}\hspace{.4pt}1109\discretionary{/}{%
}{/}TVCG\hspace{.1pt}\discretionary{.}{%
}{.}\hspace{.4pt}2017\hspace{.1pt}\discretionary{.}{%
}{.}\hspace{.4pt}2744359}}}


\bibitem{tsoi_approach_2021}
N.~Tsoi, M.~Hussein, O.~Fugikawa, J.~D. Zhao, and M.~V{\'a}zquez.
\newblock An approach to deploy interactive robotic simulators on the web for hri experiments: {{Results}} in social robot navigation.
\newblock In {\em 2021 {{IEEE}}/{{RSJ International Conference}} on {{Intelligent Robots}} and {{Systems}} ({{IROS}})}, pp. 7528--7535. IEEE, 2021. \href{https://doi.org/10.1109/IROS51168.2021.9636319}
{doi: {{%
10\hspace{.1pt}\discretionary{.}{%
}{.}\hspace{.4pt}1109\discretionary{/}{%
}{/}IROS51168\hspace{.1pt}\discretionary{.}{%
}{.}\hspace{.4pt}2021\hspace{.1pt}\discretionary{.}{%
}{.}\hspace{.4pt}9636319}}}


\bibitem{turton_etk_2017}
T.~L. Turton, A.~S. Berres, D.~H. Rogers, and J.~Ahrens.
\newblock {{ETK}}: An evaluation toolkit for visualization user studies.
\newblock In {\em Proceedings of the {{Eurographics}}/{{IEEE VGTC Conference}} on {{Visualization}}: {{Short Papers}}}, {{EuroVis}} '17, pp. 43--47. Eurographics Association, Goslar, DEU, June 2017. \href{https://doi.org/10.2312/eurovisshort.20171131}
{doi: {{%
10\hspace{.1pt}\discretionary{.}{%
}{.}\hspace{.4pt}2312\discretionary{/}{%
}{/}eurovisshort\hspace{.1pt}\discretionary{.}{%
}{.}\hspace{.4pt}20171131}}}


\bibitem{vanderplas_altair_2018}
J.~VanderPlas, B.~E. Granger, J.~Heer, D.~Moritz, K.~Wongsuphasawat, A.~Satyanarayan, E.~Lees, I.~Timofeev, B.~Welsh, and S.~Sievert.
\newblock Altair: {{Interactive Statistical Visualizations}} for {{Python}}.
\newblock {\em Journal of Open Source Software}, 3(32):1057, Dec. 2018. \href{https://doi.org/10.21105/joss.01057}
{doi: {{%
10\hspace{.1pt}\discretionary{.}{%
}{.}\hspace{.4pt}21105\discretionary{/}{%
}{/}joss\hspace{.1pt}\discretionary{.}{%
}{.}\hspace{.4pt}01057}}}


\bibitem{verbisoftware_maxqda_2018}
{VERBI Software}.
\newblock {{MAXQDA Analytics Pro}}, 2018.

\bibitem{yang_correlation_2019}
F.~Yang, L.~T. Harrison, R.~A. Rensink, S.~L. Franconeri, and R.~Chang.
\newblock Correlation {{Judgment}} and {{Visualization Features}}: {{A Comparative Study}}.
\newblock {\em IEEE Transactions on Visualization and Computer Graphics}, 25(3):1474--1488, Mar. 2019. \href{https://doi.org/10.1109/TVCG.2018.2810918}
{doi: {{%
10\hspace{.1pt}\discretionary{.}{%
}{.}\hspace{.4pt}1109\discretionary{/}{%
}{/}TVCG\hspace{.1pt}\discretionary{.}{%
}{.}\hspace{.4pt}2018\hspace{.1pt}\discretionary{.}{%
}{.}\hspace{.4pt}2810918}}}


\bibitem{yang_considering_2024}
J.~Yang, A.~M. McNutt, and L.~Battle.
\newblock Considering {{Visualization Example Galleries}}.
\newblock In {\em 2024 {{IEEE Symposium}} on {{Visual Languages}} and {{Human-Centric Computing}} ({{VL}}/{{HCC}})}, pp. 329--343, Sept. 2024. \href{https://doi.org/10.1109/VL/HCC60511.2024.00043}
{doi: {{%
10\hspace{.1pt}\discretionary{.}{%
}{.}\hspace{.4pt}1109\discretionary{/}{%
}{/}VL\discretionary{/}{%
}{/}HCC60511\hspace{.1pt}\discretionary{.}{%
}{.}\hspace{.4pt}2024\hspace{.1pt}\discretionary{.}{%
}{.}\hspace{.4pt}00043}}}


\bibitem{zacks_bars_1999}
J.~Zacks and B.~Tversky.
\newblock Bars and lines: {{A}} study of graphic communication.
\newblock {\em Memory \& Cognition}, 27(6):1073--1079, Nov. 1999. \href{https://doi.org/10.3758/BF03201236}
{doi: {{%
10\hspace{.1pt}\discretionary{.}{%
}{.}\hspace{.4pt}3758\discretionary{/}{%
}{/}BF03201236}}}


\bibitem{zhao_libra_2025}
Y.~Zhao, Y.~Wang, X.~Luo, Y.~Wang, and J.-D. Fekete.
\newblock Libra: {{An Interaction Model}} for {{Data Visualization}}.
\newblock In {\em {{SIGCHI Conference}} on {{Human Factors}} in {{Computing Systems}}}, 2025. \href{https://doi.org/10.1145/3706598.3713769}
{doi: {{%
10\hspace{.1pt}\discretionary{.}{%
}{.}\hspace{.4pt}1145\discretionary{/}{%
}{/}3706598\hspace{.1pt}\discretionary{.}{%
}{.}\hspace{.4pt}3713769}}}


\end{thebibliography}

\appendix 

\clearpage

\begin{strip} 
\noindent\begin{minipage}{\textwidth}
\makeatletter
\centering%
\sffamily\bfseries\fontsize{15}{16.5}\selectfont
\thetitle \\[.5em]
\large Appendix\\[.75em]
\makeatother
\normalfont\rmfamily\normalsize\noindent\raggedright In this appendix we provide additional tables, plots, and charts that show data beyond the material that we could include in the main paper due to space limitations or because it was not essential for explaining our approach.
\end{minipage}
\end{strip}

\appendix
\renewcommand{\thefigure}{Appendix \Alph{section}--\arabic{figure}}
\setcounter{figure}{0}

\section{Details for \jndstudytitle}
\label{sec:app-jnd}

Harrison et al.~\cite{harrison_ranking_2014} investigated how nine commonly used visualizations convey correlations \cite{harrison_ranking_2014}. 
Specifically, they examined if participant's perceived differences in correlation, measured by just-noticeable differences (JNDs), differed when using scatterplots versus other chart types. Among their findings was that  scatterplots performed best for both negatively and positively correlated data, while Parallel Coordinate Plots (PCPs) performed well with negatively correlated data, but poorly with positively correlated data. Other tested visualization techniques weren't as well suited to distinguish degrees of correlations.  

In their large-scale crowdsourced experiment with 1,687 participants they used a staircase method, i.e., increasing the difficulty of trials when the previous answer was correct and vice versa. 

We tested four visualization types in a pre-registered experiment (\url{https://osf.io/t6gp2}): Scatterplots, 
Parallel Coordinate Plots (both of which were included in the original study), Hexbin Scatterplots, and sorted 1D Heatmaps 
(see  Figures \ref{fig:jnd-distro}) for positive and negative correlations. 
We did not re-test poorly performing visualization types from the original study, such as donut charts and line charts, as the estimated JNDs were so large so as to violate the original study assumptions \cite{harrison_ranking_2014, kay_webers_2016}. 
We extended the study design, which tested two correlation values per person ($r = 0.3, 0.6$) to three per person: $r = 0.3, 0.6, 0.9$, which covers low, medium, and high correlation values. 
As in the original experiments, we ``approached'' correlation values from above and below using a staircase methodology. We employed a between-subjects design for different visualization types and correlation direction: positive vs. negative. 

We used pre-generated datasets to create the charts, which we have provided in the supplementary materials. 
We use reVISit's dynamic sequence block feature, which allows granular control of the sequence of the stimulus, particularly for dynamically deriving the stimulus parameters. 

In our 2-AFC staircase procedure, the correlations were adjusted based on participant responses. Correct answers decreased the correlation difference by $0.01$ between the base correlations, while incorrect answers increased it by $0.03$. 
This process continues until convergence based on an F-test of the last 24 trials is achieved or a participant has completed 50 trials (following, e.g. \cite{harrison_ranking_2014}).

In the study, we also introduced a dynamic and complex attention check method. Every 10th trial in our studies is an attention check, where we compare the correlation of $0.01$ and $1$. The correct answer in the attention check is reversed from the participant’s previous selection location to detect repeated clicks. As preregistered, participants failing more than 20\% of the attention checks were excluded from the analysis. 

The reVISit Analysis view provided us with a robust interface to filter participants out based on attention checks and the status of the responses. The data was exported from the reVISit Analysis view in a tidy CSV format. Out of the 240 recruited participants, 31 were excluded from the analysis due to failing more than 20\% attention checks and incomplete responses.

\subsection{Results}
\label{sec:jndresults}
We analyzed the JNDs across 8 conditions, 4 visualization types (scatterplot, parallel coordinate plot, hexbin plot, and heatmap), and two correlation directions (positive and negative. The distribution of JNDs is visualized in \figref{fig:jnd-distro}. Lower JNDs denote easier detection of correlation differences.

Scatterplots consistently performed well, which confirms the findings in the original study. Hexbin plots also performed well, but experienced some outliers especially in lower correlation. Parallel coordinate plots also performed well, especially for negative correlations. We experienced some asymmetry in positive and negative correlations which coincides with the findings in the original study. For example, at $|r| = 0.6$ in the \textit{above} condition, parallel coordinate plots had a JND of $M = 0.10,\; [0.09,\; 0.10]$ for negative correlations, but $M = 0.19,\; [0.17,\; 0.22]$ for positive correlations.

1D heatmaps produced the highest JNDs, making it the least effective among the visualization types tested for effectively conveying correlation differences. For example, at $r = +0.3$ in the \textit{above} condition, heatmaps had a JND of $M = 0.32,\; [0.27,\; 0.37]$, compared to PCPs at $M = 0.29,\; [0.27,\; 0.32]$,  $M = 0.24,\; [0.21,\; 0.28]$ for scatterplots, and $M = 0.21,\; [0.16,\; 0.27]$ for hexbin plots. Furthermore, positively correlated heatmaps had higher JNDs and greater variances compared to negatively correlated heatmaps. For instance: at $|r| = 0.3$ in the \textit{above} condition, positive heatmaps had $M = 0.32,\; [0.27,\; 0.37]$, while negative heatmaps were slightly lower at $M = 0.27,\; [0.25,\; 0.29]$.

For example, at $r = +0.3 $, in the \textit{above} condition, the highest JNDs were observed in heatmaps ($M = 0.32,\; [0.27,\; 0.37]$). In comparison, PCPs showed $M = 0.29,\; [0.27,\; 0.32]$, $M = 0.24,\; [0.21,\; 0.28]$ for scatterplots, and hexbins performed best with $M = 0.21,\; [0.16,\; 0.27]$.

Generally, we observed asymmetries between the \textit{above} and \textit{below} approaches. For example, at $r = +0.6$, scatterplots had a JND of $M = 0.12,\; [0.10,\; 0.14]$ in the \textit{above} condition, but a higher JND of $M = 0.17,\; [0.13,\; 0.21]$ in the below condition. Also, JNDs increased as the base correlation decreased, which is consistent with the original study. In the above condition, scatterplots had a JND of $M = 0.24,\; [0.21,\; 0.28]$ at $r = +0.3$, which decreased to $M = 0.12,\; [0.10,\; 0.14]$ at $r = +0.6$, and further to $M = 0.05,\; [0.04,\; 0.05]$ at $r = +0.9$.

\subsection{Discussion}
 
Our results are consistent with the findings reported in the original study by Harrison et al.~\cite{harrison_ranking_2014}, particularly the performance of scatterplots and parallel coordinate plots (see Section~\ref{sec:jndresults}). We also found that JND values decreased as the base correlation increased, which is consistent with Weber's law, another trend reported in the original study ~\cite{harrison_ranking_2014}.  This suggests that visual encodings should be carefully chosen when showing weak correlations, which are more susceptible to mistakes.

Additionally, we generally observed tighter confidence intervals across conditions. This may be attributable to improved data quality control measures in our study design. Specifically, we incorporated attention checks throughout the experiment and excluded a total of 31 participants from the analysis due to high attention check failure and incomplete submissions. These measures likely contributed to reduced variability and tighter intervals.

For hexbin plots, we used a larger dataset size ($n = 1000$) compared to the other visualization types ($n = 100$) to better match real-world usage of hexbin plots, which are commonly used when scatterplots are not scalable enough. Our results show that hexbin plots can successfully convey correlations, especially for large datasets. In some conditions, hexbin plots have performed the best, even yielding lower mean JNDs than the scatterplots, which performed best in the original study. We highlight one such instance in Section~\ref{sec:jndresults}.

The relatively poor performance of heatmaps in our study suggests that while they are commonly used in practice for visualizing matrices or large-scale patterns, they may not be well-suited for tasks that require precise judgment of correlations. Furthermore, design-specific factors such as the pairwise comparison format, color encoding or layout density, etc. may also contribute to the lower performances. Further study is needed to understand how design choices affect their perceptual of correlations. 

Overall, our inclusion of hexbin and heatmap visualizations extends the design space explored in prior work. Hexbin plots show strong potential for large datasets and lower correlations. Heatmaps, despite their lower performance, may still be useful for compact layouts.

The stimulus was implemented using React and TypeScript. 

The study interface and data are viewable in dissemination mode at \url{https://revisit.dev/replication-studies/}. The data generation and study specification reVISitPy script, as well as all the analysis code is available at \url{https://github.com/revisit-studies/replication-studies-analysis}.

The experiment was reviewed by the University of Utah IRB and deemed exempt from full board review (IRB 00187851). Participants were paid a fixed amount per study, with a median completion time ranging from $13:26$ to $21:01$ minutes, and an average hourly compensation ranging from $\$11.42$ to $\$17.88$, as reported by Prolific. First, we held an internal pre-pilot run with 3 participants. Next, we did an external pilot with 16 crowdsourced participants, with 2 participants in each condition on prolific. In our final study, we employed 240 crowdsourced participants with 30 participants per condition across 8 studies.


\section{Details for \patternstudytitle}
\label{sec:pattern-study-details}

\begin{figure*}[t]
    \centering
    \includegraphics[width=0.9\linewidth]{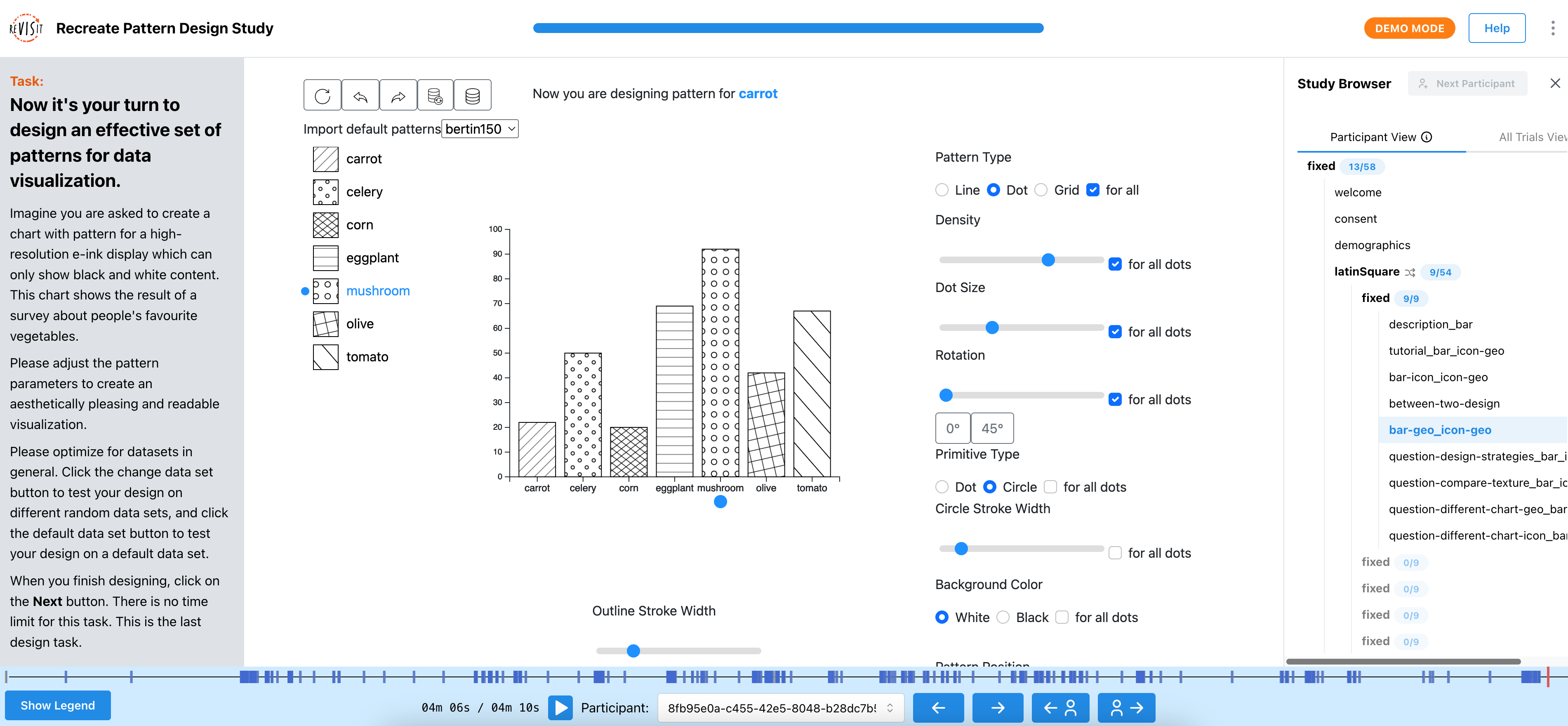}
    \caption{Participant result for pattern-designs using a strategy based on leveraging default patterns. [\ourhref{https://revisit.dev/replication-studies/pattern-design-study/V3ZPZStsZmxBWDNta2p4MyttU1dJZz09?participantId=8fb95e0a-c455-42e5-8048-b28dc7b56230}{View Result}].}
    \label{fig:pattern-example}
\end{figure*}

Here we provide details on our replication of He et al.'s study \cite{he_design_2024} on designing black-and-white patterns for visualization, sketched in \autoref{sec:pattern-study}. In their study, they asked 30 visualization designers to create two types of black-and-white patterns (geometric and iconic) for a given chart type (bar, pie, or map) using their online pattern design tool, which allowed participants to adjust a wide range of parameters. 
After completing their two designs, participants were shown their own patterns and asked to describe their design goals and strategies. They were also asked to compare their geometric and iconic designs. In addition, their pattern designs were applied to two other chart types (e.g., if a participant is in a bar chart, their patterns were applied to pie and map charts), and participants were asked whether the patterns still work.
The authors then coded participants' responses and categorized the goals into those related to readability and aesthetics, and summarizing corresponding design strategies. They also discussed experts' opinions on using geometric and iconic patterns in visualizations.

We chose to replicate this study to show that reVISit 
(a) supports a mixed design method (between-subject design on chart type and within-subject design on pattern order) and (b) enables researchers to display answers (in this case, pattern designs) from earlier tasks in later ones. The original study was developed using FROE~\cite{jansen_froe_2025} where this functionality had to be manually implemented by the designer. In addition, we also show that (c) existing online tools can be effectively integrated into the reVISit framework. In contrast to the previous study, we fully instrumented the stimulus so we could analyze design strategies and reason about the number of interactions associated with good or bad designs.  Both, integration of existing tools and instrumentation, are important for studies on visual representation design and the evaluation of authoring tools. 

To replicate this study, we generally follow the original study design with minor changes. We pre-registered our study on OSF (\url{https://osf.io/w6csx}). The study was reviewed by the University of Utah IRB (IRB number: 00188052) and deemed exempt from full board review. 

The original study was implemented with vanilla JavaScript and PHP. We adapted their description, pattern design tool and post-questionnaires as HTML stimuli, other pages as markdown stimuli in reVISit. For pages that need to show different content based on different conditions (e.g., the description page), we pass the parameters from the study configuration to the components. For the pages that need to show content based on previous answers (e.g., the design comparison page), we take advantage of reVISit's stepped function to pass the answers. 

We refined the UI design with the support of reVISit UI configuration, for example, by using the side-bar to avoid scrolling between instructions and stimuli. 

To better understand pattern design, it's helpful to capture the design process itself---which is an aspect missing from the original study. In the original study, for the design strategies, the authors only analysis the text responses. With reVISit and Trrack \cite{cutler_trrack_2020}, we fully record participants' interaction provenance, and we can review and analyze them later using the framework’s replay interface. This leads us to investigate a new research question: Does the engagement in the interactions affect the final design results?


\subsection{Results}
We received 20 valid responses (9\texttimes{} bar, 5\texttimes{} pie, 6\texttimes{} map) from design experts (prior experience in visualization design: mean = 9.5 years, SD = 12.5 years). 

We coded participants' free-text responses based on the codebook of the original study. Participants stated similar design goals as the original study, they mainly focus on  distinguishability (15 \texttimes{}), visual clarity (5 \texttimes{}), semantic association (5 \texttimes{}), visual pleasure (7 \texttimes{}), and visual balance (3 \texttimes{}). They also shared similar design strategies and opinions as reported in the original study. Two new interesting design strategies emerged: When designing geometric patterns, one participants attempted to create new shapes by playing with the basic shapes to create new shapes, and another designer experimented with dot density to create different shades of gray, akin to the halftoning technique.

Based on the tracked provenance data, we counted the number of interaction events associated with each design. In addition, one researcher rated each design on a 1-7 scale for aesthetics (with the 5-item BeauVis scale \cite{he_beauvis_2023}) and one question for perceived vibratory effect. 
We also qualitatively analyzed the participants' interaction behaviors during the design process by using the reVISit replay interface.

As shown in \ref{fig:pattern-beauvis-events}, we did not observe clear correlations between the number of interaction events and the final design ratings for either geometric or iconic patterns. This may be due to the influence of different design workflow and the subjective nature of our rating. 
Some participants produced highly rated designs despite generating few interaction events. In the case of geometric patterns, this was often because they made only minor adjustments to one of the default pattern sets 
(as in \ourhref{https://revisit.dev/replication-studies/pattern-design-study/MTdrZWtXYTNtR1hvUU9zNit1Z1pRQT09?participantId=dce48040-7493-4d9f-a681-2a5726962ef2}{this} or \ourhref{https://revisit.dev/replication-studies/pattern-design-study/MTdrZWtXYTNtR1hvUU9zNit1Z1pRQT09?participantId=e5da64b7-4415-4296-93e5-daa52ba45d7b}{that} participant)

Through the reVISit replay interface, we noticed some common strategies that designers use. 
A typical workflow involved an initial exploration of different default pattern sets, followed by iterative refinement of individual patterns, and finally test and refine design with different datasets (see Figure~\ref{fig:pattern-example}). This workflow is natural and effective. 
However, we also observed two designers who achieved high-quality results despite deviating from this workflow. These participants began by editing patterns directly and did not evaluate their designs across datasets (\ourhref{https://revisit.dev/replication-studies/pattern-design-study/V3ZPZStsZmxBWDNta2p4MyttU1dJZz09?participantId=6d78c17f-d258-4796-bc3f-bc511b4ff87c}{for example}). Their behavior may reflect their personal design habits or may have been influenced by the separation between pattern parameter controllers and dataset controllers on the interface. For iconic patterns, a commonly observed behavior was testing different icon styles (see \autoref{fig:texture-stim}).
This preference may be because icon style controllers are the top row of all controllers, or because icon styles have a clear visual impact on overall design compared to other parameters. Another effective design strategy for iconic patterns we noticed is to the use of the ``for all'' button to simultaneously modify parameters across all pattern categories, which can produce a balanced design (like \ourhref{https://revisit.dev/replication-studies/pattern-design-study/MTdrZWtXYTNtR1hvUU9zNit1Z1pRQT09?participantId=6d78c17f-d258-4796-bc3f-bc511b4ff87c}{this} or \ourhref{https://revisit.dev/replication-studies/pattern-design-study/MTdrZWtXYTNtR1hvUU9zNit1Z1pRQT09?participantId=8fb95e0a-c455-42e5-8048-b28dc7b56230 }{that}).
None of the participants mentioned these interaction strategies in their text responses. We were only able to notice them using the replay, highlighting the platform's value in uncovering subtle yet impactful design behaviors. These observations can inspire further investigation on the impact of different design strategies or interface layout of the design tool.

\section{Details for Replication of Search in Visualization Study}
\label{sec:search-study-details}

In the study with Feng \etal{} \cite{feng_effects_2018}, three different visualizations were studied with and without the presence of search functionality. 
In an online study with 830 online participants over 5 different experiments, participants were first provided a tutorial of the interface, followed by the visualization itself where they are asked to analyze the data. 
After their exploration, they were asked to provide their takeaways and findings through text-based inputs. 
The authors derived quantitative and qualitative measures from collected data including intent, active search count, exploration time, average visit time, etc. 
Results showed that participants who used search engaged with data items highlighted during search for longer periods of time.
Results also showed that people using search explored a wider range of values compared to those who did not use search.

We chose to replicate this study because it (a) utilizes interaction logging and (b) incorporates a text-based think-aloud protocol, though the original implementation relied on text-based input recorded after the exploration task. 
Our replication closely follows the original study design, focusing on two of the three visualizations--a bubble chart of colleges and The New York Times' ``255 Charts''--with minor modifications.

Since reVISit integrates the Trrack library, which is a more systematic way to log interactions, this study aligns well with our replication goals. 
Additionally, rather than relying exclusively on user interactions and post-task textual responses about what, why, and how participants searched for items, incorporating a think-aloud capability could provide deeper insights into their thought processes.
We pre-registered our study on OSF (\url{https://osf.io/7baqy}) with minor changes from the original study:
First, one condition (Colleges Bubble Chart) was re-implemented using React. 
The other condition (255 charts), was kept in its original form as much as possible, and displayed via a website component. 

\textbf{Bubble Chart}: In our study with bubble chart, we used the same 300 colleges from the original study, where each college was represented by a circle. 
The radius, color, and distance from the center encoded the college's annual cost,
median student earnings, 
and admission rate, respectively. 
Hovering over a circle revealed a detailed view with text values for the underlying data.
For participants in the search-enabled condition, a search box was positioned in the top-left corner of the bubble chart.
Searches and highlights were triggered dynamically as characters were typed. 
To display search results, selected data items retained their full opacity,
while unselected items were slightly faded to reduce visual prominence.

\begin{figure*}[ht]
    \centering
    \includegraphics[width=0.9\linewidth]{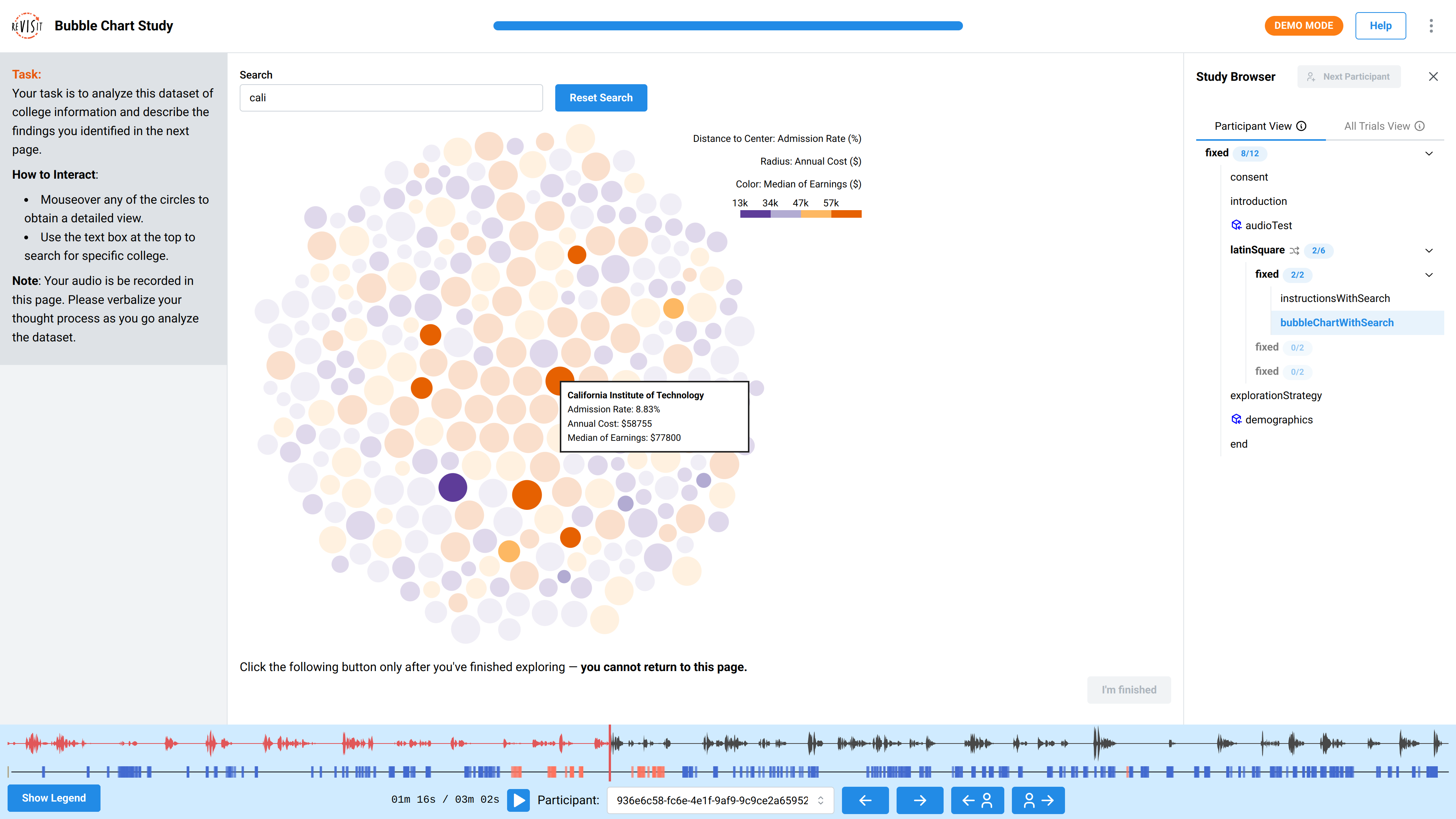}
    \caption{An example replay of a participant exploring the bubble chart study where they are using the search box. 
    [\ourhref{https://revisit.dev/replication-studies/bubblechart-study/LzE2MTl4ZVRMTk5nSFlNYmd1ZDhjZz09?participantId=936e6c58-fc6e-4e1f-9af9-9c9ce2a65952}{View Result}].
    }
    \label{fig:bubble-search-example}
\end{figure*}

\begin{figure*}[t]
    \centering
    \includegraphics[width=0.9\linewidth]{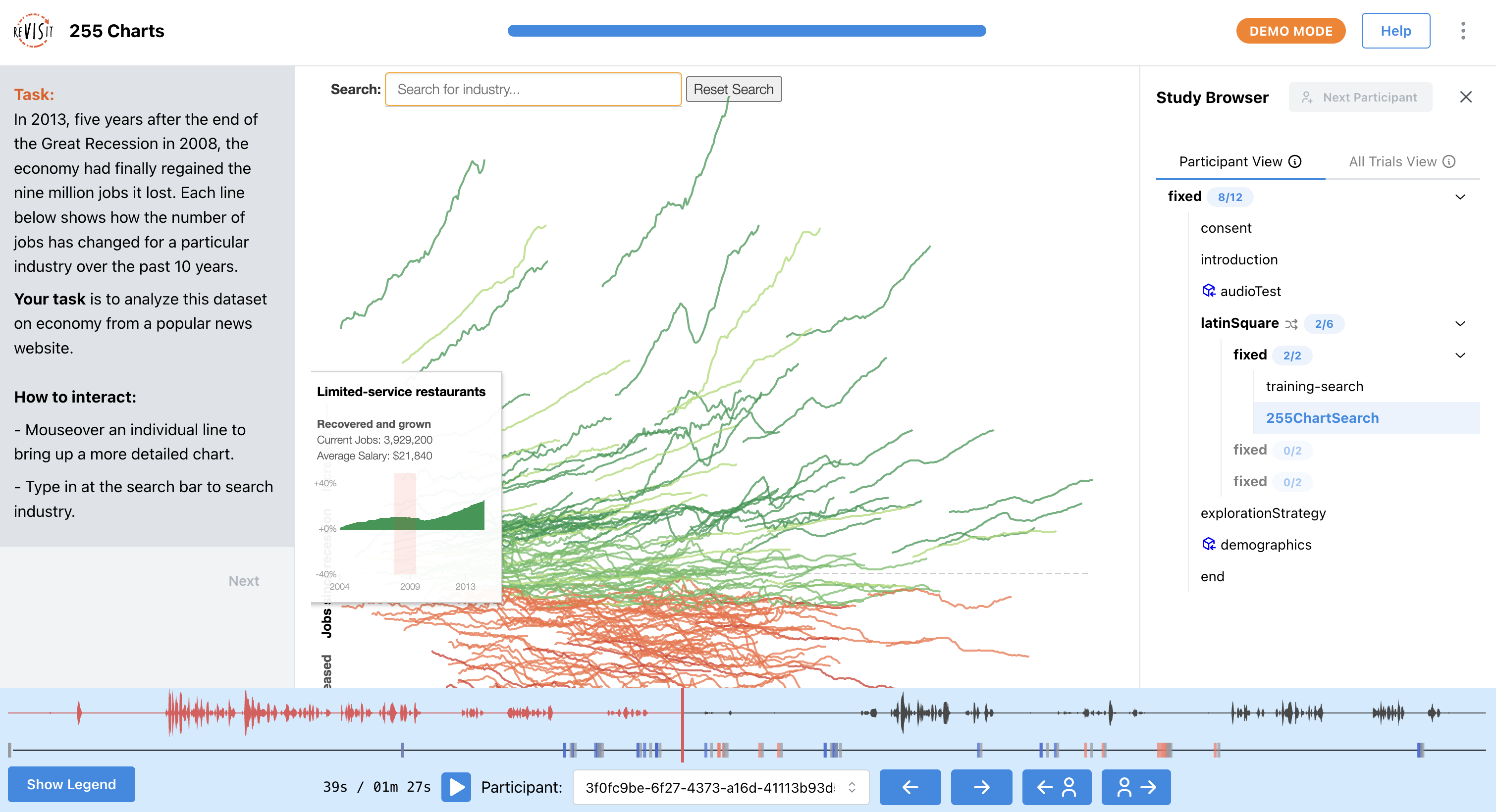}
    \caption{A replay of a participant who took the 255-charts study. The blue bar indicates hover event and the orange bar indicates search activities (This information can be checked by clicking "Show Legend" button)}
    \label{fig:255-example}
\end{figure*}

\textbf{255 Chart}: In our other study with 255 charts, we replicated visualization from NYTimes, titled ``How the Recession Shaped the Economy, in 255 Charts''. It included 255 line charts distributed across the viewport in a scatterplot-like fashion. 
Each line represents how a particular industry of the US Economy, such as Home Health Care Service, grew or declined from 2004 to 2014. 
On mouse hover over an industry, an animated tooltip reveals a more detailed line-chart view with specific values, years and industry information.
In both cases, the search box in the search-enabled condition is displayed in the top left of the screen, allowing users to search for college or industry names. Fig. \ref{fig:255-example} shows an example of the replay of a participant.

\subsection{Results}
\label{sec:search-results}
99 participants were recruited for Bubble visualization, of which one participant did not interact with any elements for more than 500ms.
This resulted in 64 participants in the search condition and 34 in the non-search condition.
Similarly, for the 255 charts, we recruited 100 participants. However, 7 participants did not interact with any element for more than 500ms. This resulted in 62 completed trials in the search condition and 31 in the non-search condition. 

Quantitative results largely align with the prior study. In particular, we find that participants generally explored a similar number of items regardless of the presence of search (\autoref{fig:search-distribution}). However, as in the prior study, we did find that, when participants engaged with items that were highlighted during search, they interacted with them for significantly longer.
Specifically, a Wilcoxon test on item engagement durations during search versus outside of search were statistically significant for both visualization types, with Bubble $W=1451, p=2.4384e-06, d=0.67 ~ [0.24, 1.11]$ and 255 $W=352, p=0.00918, d=0.72~[0.11, 1.33]$.

For qualitative analysis, we used the OpenAI Whisper model to generate transcripts for all participants from the think-aloud audio recording. 
Then we used LLooM \cite{lam_concept_2024} to infer concepts from the transcripts to further guide analysis.
We then conducted a qualitative review by exploring individual participants in the reVISit replay interface (\eg~\autoref{fig:pattern-example}).

One of the findings from the original paper is that participants reported intentionally seeking specific data items within the visualization. 
For example, one participant searched for several industries they were interested in. 
When they looked up computer-related industries, they quickly identified that Computer Systems Design and Programming showed increasing wages, while Computer and Peripheral Equipment Manufacturing showed decreasing wages—an observation that surprised them.(\ourhref{https://revisit.dev/replication-studies/255chart-study/LzE2MTl4ZVRMTk5nSFlNYmd1ZDhjZz09?participantId=3a12c79b-75ce-4c1a-82ed-8b95f8523ae7}{Replay})
We observe a similar pattern in the bubble chart study. In this example: (\ourhref{https://revisit.dev/replication-studies/bubblechart-study/LzE2MTl4ZVRMTk5nSFlNYmd1ZDhjZz09?participantId=936e6c58-fc6e-4e1f-9af9-9c9ce2a65952}{Replay}), a participant begins by exploring colleges they are familiar with but quickly notices that the data challenges their prior beliefs. This discrepancy invites them to further explore neighboring and more central colleges within the bubble chart.

With easier-to-build provenance tracking, activity replay interfaces, and think-aloud study provided by reVISit, we now have a more powerful tool to explore participants' behavior by examining their interactions and their provided exploration thoughts.
Many previous studies rely solely on participants’ responses to predefined questionnaires. With reVISit, studies such as this can observe more deeply how participants interact with visualizations, and learn more about what participants are thinking about through the process.

In conducting qualitative analyses with the provenance tracking and think-aloud data, we discovered new patterns in how people interact with the provided visualizations. 
For example, many users began exploring the 255 charts example by examining maximum and minimum values. 
As a result, they tend to hover over the edges and borders before exploring industries in the center of the chart. 
These behaviors can be verified through both interaction replays and think-aloud transcripts.
While this behavior was hypothesized from aggregate data in the original study, the reVISit toolkit lowers the barrier to deeper-grained analysis, making these types of analyses more accessible and feasible for research teams.

These capabilities are not without limitations and considerations, however.
For example, it is possible that asking users to think-aloud while also engaging with an exploratory visualization could lead to interruptions in their natural analysis flow.
While our analyses still indicated significant differences in exploration time for elements that participants searched for, such concerns are still potentially valid.
At the same time, easier access to time-stamped exploration and voice data may make it possible to construct new robust metrics for peoples' engagement with visualization stimuli--- which could be a fruitful area of future work.

\section{Additional Figures showing reVISit and reVISitPy}

\begin{figure*}[!h]
    \centering
    \includegraphics[width=0.8\linewidth]{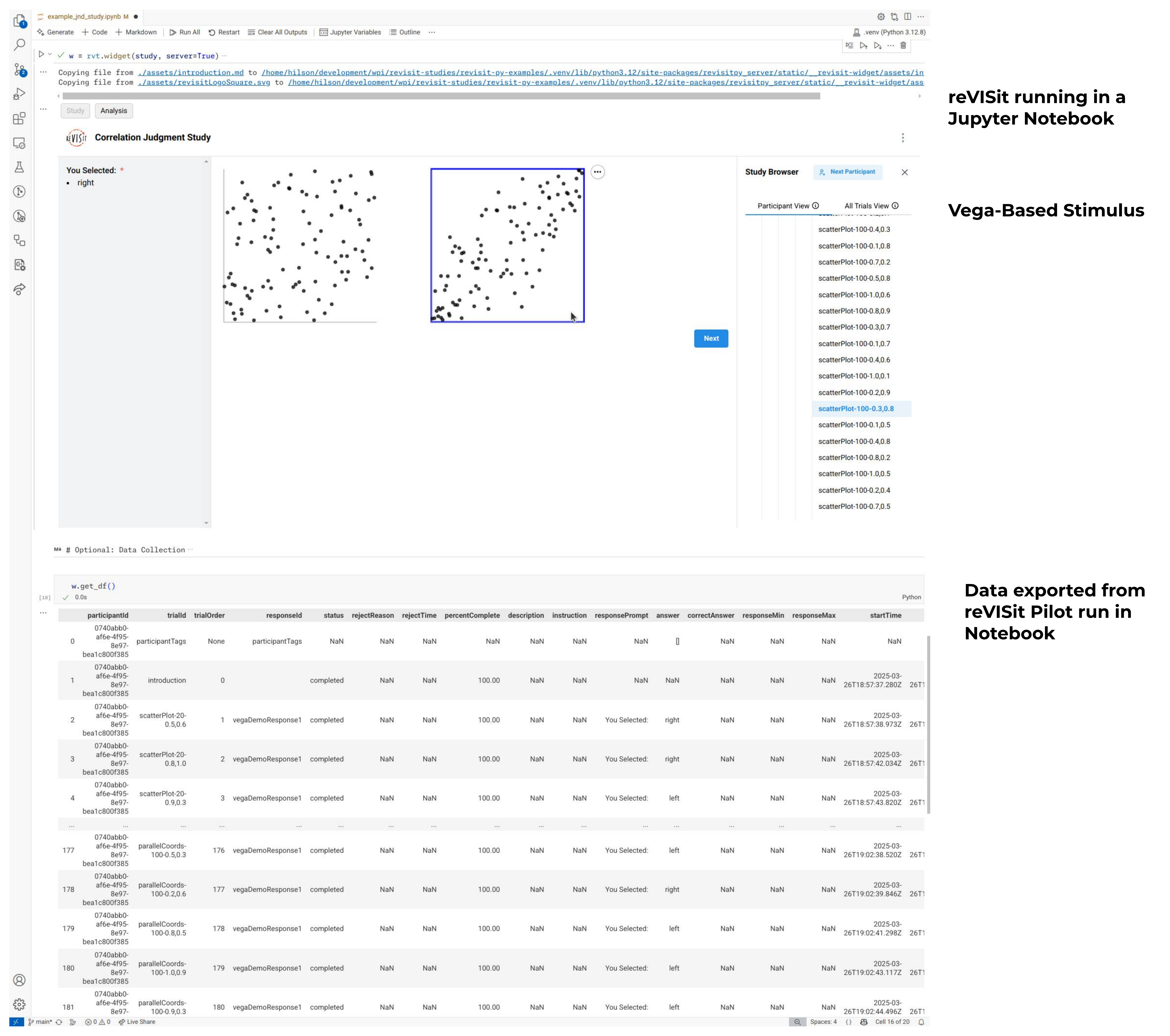}
    \caption{Example of how reVISitPy can be used to prototype a study from a single Jupyter Notebook, from study specification, stimulus design, debugging, piloting, and preliminary data analysis. Full source: \url{https://revisit.dev/docs/revisitpy/examples/example_jnd_study/}}
    \label{fig:revisitpy}
\end{figure*}

\autoref{fig:revisitpy} shows an example of how reVISitPy can be used to prototype a study within a Jupyter notebook. 

\begin{figure*}[!h]
    \centering
    \includegraphics[width=\linewidth]{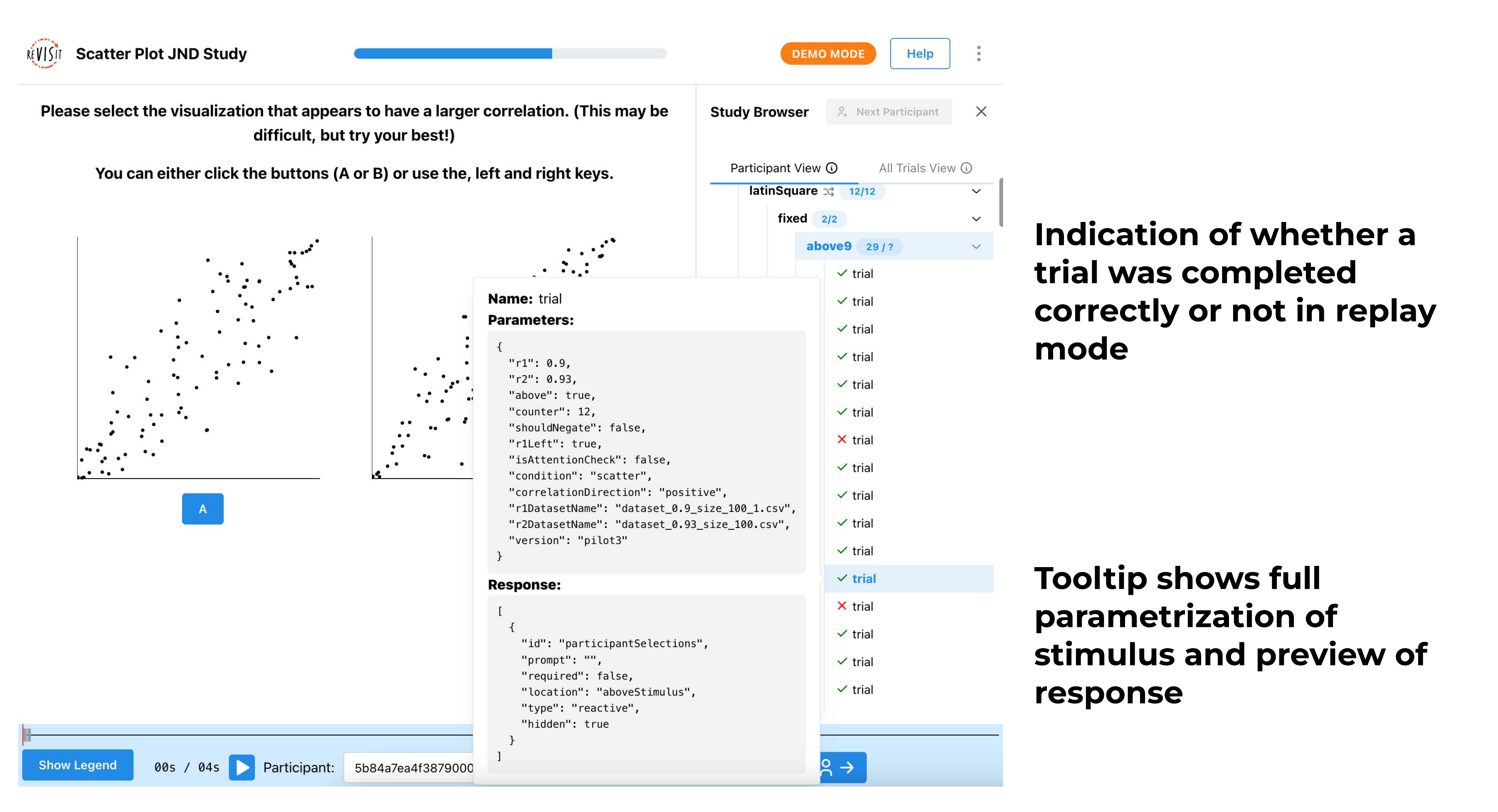}
    \caption{In replay mode (i.e., when viewing a participants completed tasks), the study browser indicates whether a trial was answered correctly or not. Hovering over a trial also shows the configuration of the trial. \href{https://revisit.dev/replication-studies/ScatterJND-study/WHUwRzhDeHVFMUh5b2VGSzNFVDhMQT09/UGczdnZrNnV5a1d4aitzaCtCNVRlZz09?participantId=5b84a7ea4f38790001778d53}{Source}}
    \label{fig:study-browser-replay}
\end{figure*}

\autoref{fig:study-browser-replay} shows how additional information is displayed in the study browser when replaying a study. 

\end{document}